# Consistent continuous defect theory

Ali R. Hadjesfandiari, Gary F. Dargush

*Department of Mechanical and Aerospace Engineering*
*University at Buffalo, The State University of New York, Buffalo, NY 14260 USA*

ah@buffalo.edu, gdargush@buffalo.edu

August 15, 2018

**Abstract**

By investigating the benefits and shortcomings of the existing form of continuous defect theory (CDT) and using recent advancements in size-dependent continuum mechanics, we develop a fully coherent theoretical framework, denoted as Consistent Continuous Defect Theory (C-CDT). Among several important potential applications, C-CDT may provide a proper foundation to study the continuum theory of crystal plasticity. The development presented here includes an examination of the character of the bend-twist tensor, Weingarten's theorem, Burgers and Frank vectors, continuous dislocation and disclination density tensors, and the dualism between geometry and statics of CDT based on couple stress theory (CST). Then, by using Consistent Couple Stress Theory (C-CST), the new C-CDT is derived in a totally systematic manner. In this development, the geometry of C-CDT is dual to the statics formulation in C-CST. Previously, the fundamental step in the creation of C-CST was recognizing the skew-symmetric character of the couple-stress tensor, which requires the skew-symmetrical part of the bend-twist tensor to be the additional measure of deformation in size-dependent continuum mechanics. Via Weingarten's theorem and arguments from C-CST, we establish that in defect theory the dislocation density tensor must be skew-symmetric and thus can be represented by an equivalent dislocation density vector. In addition, we investigate the character of a classical version of C-CDT with unexpected consequences. For full consistency, there can be no continuous dislocation density tensor within classical continuum mechanics, and the continuous disclination density tensor becomes symmetric. This clearly is analogous to the absence of couple-stresses and the symmetry of force-stresses in classical continuum mechanics.

# 1. Introduction

With the motivation from Weingarten's theorem (Weingarten, 1901), Volterra introduced two sets of fundamental discrete defects: dislocations and disclinations in multiply connected bodies (Volterra, 1907). These discrete defects are measured by the Burgers vector and Frank vector, which represent the discontinuity or jump in displacement and rotation vectors in a multiply connected body, respectively. In an atomistic representation of crystal plasticity, these discrete dislocations and disclinations are considered to represent the plastic crystal defects that arise from translational and rotational lattice incompatibility, respectively (Love, 1920; Frank, 1958; Kröner and Anthony, 1975). However, in a continuous defect theory, the internal defects must be expressed by continuous tensor quantities, such as dislocation and disclination density tensors (Kröner, 1958, 1968, 2001; Nabarro, 1967; Schaefer, 1968, 1967a,b; Anthony et al. 1968; Anthony, 1970; deWit, 1960, 1970, 1973a-c). Of fundamental importance is Weingarten's theorem, which provides the means to define disclination and dislocation density tensors from incompatible elastic or plastic deformation (Nabarro, 1967, Schaefer, 1968, 1967a,b; Anthony, 1970; deWit, 1970, 1973a-c).

The work of Kondo (1968), in reducing the study of many physical phenomena to a study of geometry, has given the notion that there is a duality between the geometry of defects and the statics of continuum mechanics. In particular, Anthony et al. (1968) have presented a continuous defect theory based on the dualism between geometry and statics within Cosserat continuum mechanics (Cosserat and Cosserat, 1909), equipped with an independent rotational field. However, Weingarten's theorem indicates that this dualism should be between the geometry of continuous defect theory and the statics of couple stress theory (CST), where the rotational degrees-of-freedom are not independent, but rather are derived directly from the curl of the displacement field. A few years later, Anthony (1970) and deWit (1970) have demonstrated this dualism between their continuous defect theory and the original couple stress theory developed by Mindlin and Tiersten (1962) and Koiter (1964). Unfortunately, this form of couple stress theory, called Mindlin-Tiersten-Koiter couple stress theory (MTK-CST), suffers from several inconsistencies (Eringen, 1968; Hadjesfandiari and Dargush, 2011, 2015a,b, 2016; Hadjesfandiari et al., 2015). For example, the couple-stress tensor in this theory is indeterminate, because it is energetically conjugate to the gradient of rotation or the full non-symmetric bend-twist-tensor.

Therefore, one suspects that the dual continuous defect theory (CDT) of Anthony (1970) and deWit (1970) might also inherit inconsistencies. Most noticeably, the appearance of the continuous climbing edge dislocation in this CDT is problematic, because it requires separation of slip planes in crystals. Clearly, this cannot be accepted in a continuum model with compatible deformation. Remarkably, the inconsistencies of this CDT are quite similar to those identified in the Mindlin-Tiersten-Koiter couple stress theory (MTK-CST). Yet, the CDT of Anthony (1970) and deWit (1970) still plays a fundamental role in developing a consistent version of continuous defect theory, as will be demonstrated here. For historical reasons, in the present paper, we try to be consistent with the definitions and notations used in the original work by deWit (1970, 1973a-c), as much as possible.

Hadjesfandiari and Dargush (2011) and Hadjesfandiari et al. (2015) have developed Consistent Couple Stress Theory (C-CST), which resolves all inconsistencies in the original Mindlin-Tiersten-Koiter couple stress theory (MTK-CST). The fundamental character of this theory is that the couple-stress tensor is skew-symmetric and is energetically conjugate to the skew-symmetric part of the bend-twist-tensor. C-CST provides a fundamental basis for the development of size-dependent theories in many multi-physics disciplines that may govern the behavior of matter at the smallest scales for which continuum theory may apply. For example, Hadjesfandiari (2013, 2014) has developed size-dependent piezoelectricity and thermoelasticity, respectively. As we shall see, C-CST also provides the fundamental basis for properly completing the important, but nevertheless imperfect, continuous defect theory of Anthony (1970) and deWit (1970).

Here, we develop C-CDT by defining consistent disclination and dislocation density tensors. The fundamental character of this theory is that the dislocation tensor is skew-symmetric and can be represented by a vector. This result also confirms that there is a dualism between the geometry of C-CDT and the formulation of statics within C-CST. Although we consider elastic-plastic material in this development, we never need to introduce any constitutive relations. This means there is no constraint on the elastic or plastic deformation. For example, the material can be considered linear, non-linear, isotropic or anisotropic in the elastic or plastic ranges. Therefore, C-CDT may provide a fundamental framework to study multi-scale crystal plasticity from a continuum mechanics perspective. Equipped with this theory, many problems, such as deforming olivine-rich rocks in the mantle (Cordier et al., 2014), can be revisited. It also should be emphasized that although discrete defect theory gives the impetus for developing C-CDT with continuous disclination and dislocation density tensors, we should not expect these theories to be completely analogous.

After developing C-CDT, we also investigate the character of classical continuous defect theory (CC-CDT) as a special case with unanticipated consequences. It is seen that the geometry of this classical defect theory with symmetric disclination density tensor is dual to its statics in classical continuum mechanics with a symmetric force-stress tensor. As a result, the complete C-CDT with dislocation and disclination densities extends classical theory containing disclinations, not dislocations. In this way, C-CDT reduces to a proper classical defect theory when the couple-stresses, and consequently, the dislocations vanish. This means for having an internal continuous dislocation density tensor in the body, the existence of couple-stresses is necessary.

The remainder of the paper is organized as follows. In Section 2, we provide the notational convention, and an overview of necessary tensor analysis. Section 3 contains a brief review of the kinematics of continua, measures of deformations and compatibility conditions. Afterwards, in Section 4, we present discrete defect theory. This includes Weingarten's theorem and Volterra's disclinations and dislocations. In Section 5, we review couple stress theory (CST), including the original indeterminate Mindlin-Tiersten-Koiter couple stress theory (MTK-CST) and C-CST, upon which the current work is based. In Section 6, we present the original continuous defect theory (CDT). In Section 7, we present this theory from a fresh perspective by introducing new concepts, and investigate its inconsistencies. Then, with all of this background firmly in place, we develop C-CDT in detail in Section 8. Afterwards, in Section 9, we investigate the character of the classical version of C-CDT, namely CC-CDT. Finally, we offer some conclusions in Section 10.

## 2. Preliminaries

In this paper, we use a combination of dyadic (symbolic) and indicial notations, because the existing literature on defect theory and couple stress theory tends to employ these two different systems. Let us consider the three dimensional orthogonal coordinate system $x_1x_2x_3$ with origin $O$ as the reference frame, where $\mathbf{i}_1$, $\mathbf{i}_2$ and $\mathbf{i}_3$ are unit base vectors. This is the main reference frame coordinate system, which we use to represent the components of fundamental tensors and tensor equations. Here a bold symbol denotes a tensor. For example, the first order tensor (vector) $\mathbf{w}$ with components $w_i$ is

$$\mathbf{w} = w_i \mathbf{i}_i \tag{1}$$

and the second order tensor $\mathbf{T}$ with components $T_{ij}$ is

$$\mathbf{T} = T_{ij} \mathbf{i}_i \mathbf{i}_j \tag{2}$$

The second order identity tensor is denoted by $\mathbf{I} = I_{ij} \mathbf{i}_i \mathbf{i}_j$, where its components are represented by Kronecker delta $\delta_{ij}$, that is

$$\mathbf{I} = I_{ij} \mathbf{i}_i \mathbf{i}_j = \delta_{ij} \mathbf{i}_i \mathbf{i}_j \tag{3}$$

The transpose of the second order tensor $\mathbf{T}$ with components $T_{ij}$ is the tensor $\mathbf{S} = \mathbf{T}^t$ with components $T_{ji}$, that is

$$\mathbf{S} = \mathbf{T}^t \qquad S_{ij} = T_{ji} \tag{4}$$

A symmetric second order tensor $\mathbf{U} = U_{ij} \mathbf{i}_1 \mathbf{i}_2$ is a tensor with

$$\mathbf{U}^t = \mathbf{U} \qquad U_{ji} = U_{ij} \tag{5}$$

while a skew-symmetric tensor $\mathbf{V} = V_{ij} \mathbf{i}_1 \mathbf{i}_2$ is a tensor with

$$\mathbf{V}^t = -\mathbf{V} \qquad V_{ji} = -V_{ij} \tag{6}$$

This skew-symmetric tensor is specified by three independent components as

$$\left[ V_{ij} \right] = \begin{bmatrix} 0 & V_{12} & V_{13} \\ -V_{12} & 0 & V_{23} \\ -V_{13} & -V_{23} & 0 \end{bmatrix} \tag{7}$$

Interestingly, this tensor can be represented by its dual vector $\vec{V}$ as

$$\vec{V} = \frac{1}{2}\boldsymbol{\varepsilon} : \mathbf{V} \qquad\qquad V_i = \frac{1}{2}\varepsilon_{ijk} V_{jk} \qquad (8)$$

where $\boldsymbol{\varepsilon}$ is the third order permutation or Levi-Civita pseudo tensor. Therefore, the components of the dual vector $\vec{V}$ are

$$V_1 = V_{23} \qquad\qquad V_2 = -V_{13} \qquad\qquad V_3 = V_{12} \qquad (9)$$

Here the arrow has been superposed to represent the vector $\vec{V}$ dual to the skew-symmetric second order tensor $\mathbf{V}$. We also notice the relation

$$\mathbf{V} = \boldsymbol{\varepsilon} \bullet \vec{V} \qquad\qquad V_{ij} = \varepsilon_{ijk} V_k \qquad (10)$$

Two useful relations between the Kronecker delta and the permutation tensor are

$$\varepsilon_{ijk}\varepsilon_{ipq} = \delta_{jp}\delta_{kq} - \delta_{jq}\delta_{kp} \qquad (11)$$

$$\varepsilon_{ijk}\varepsilon_{ijp} = 2\delta_{kp} \qquad (12)$$

The gradient of a vector $\mathbf{w}$ from left and right, respectively, are defined as second order tensors

$$\nabla \mathbf{w} = w_{j,i} \mathbf{i}_i \mathbf{i}_j \qquad (13)$$

$$\mathbf{w}\nabla = w_{i,j} \mathbf{i}_i \mathbf{i}_j \qquad (14)$$

It should be noticed that these tensors are the transpose of each other, that is $(\nabla \mathbf{w})^t = \mathbf{w}\nabla$.

The divergence of a second order tensor $\mathbf{T}$ from left and right, respectively, are defined as vectors

$$\nabla \bullet \mathbf{T} = T_{ji,j} \mathbf{i}_i \qquad (15)$$

$$\mathbf{T} \bullet \nabla = T_{ij,j} \mathbf{i}_i \qquad (16)$$

while the curl of a second order tensor $\mathbf{T}$ from left and right are

$$\nabla \times \mathbf{T} = \varepsilon_{ikl} T_{lj,k} \mathbf{i}_i \mathbf{i}_j \qquad (17)$$

$$\mathbf{T} \times \nabla = -\varepsilon_{jkl} T_{il,k} \mathbf{i}_i \mathbf{i}_j$$
$$= \varepsilon_{jkl} T_{ik,l} \mathbf{i}_i \mathbf{i}_j \qquad (18)$$

respectively. Interestingly, there is the relation

$$(\nabla \times \mathbf{T})^t = -\mathbf{T}^t \times \nabla \qquad (19)$$

Meanwhile, the curl of the skew-symmetric tensor $\mathbf{V} = V_{ij}\mathbf{i}_1\mathbf{i}_2$ can be expressed as

$$\nabla \times \mathbf{V} = \boldsymbol{\varepsilon} : \left(\nabla \vec{\mathbf{V}} \bullet \boldsymbol{\varepsilon}\right) \qquad \varepsilon_{ikl}V_{lj,k} = \varepsilon_{ikl}\varepsilon_{ljm}V_{m,k} \qquad (20)$$

By using the epsilon-delta relation (11), this can be written as

$$\nabla \times \mathbf{V} = \left(\nabla \bullet \vec{\mathbf{V}}\right)\mathbf{I} - \vec{\mathbf{V}}\nabla \qquad \varepsilon_{ikl}V_{lj,k} = V_{k,k}\delta_{ij} - V_{i,j} \qquad (21)$$

Relation (21) plays an important role in this paper.

## 3. Kinematics, measures of deformation and compatibility conditions

Consider a material continuum occupying a volume $V_0$ bounded by a surface $S_0$. The deformation of the body is represented by the continuous displacement field $\mathbf{u}$. In infinitesimal deformation theory, the displacement vector field $\mathbf{u}$ is sufficiently small that the relevant kinematical quantities are defined from its gradient represented by the distortion tensor

$$\boldsymbol{\beta} = \nabla \mathbf{u} \qquad \beta_{ij} = u_{j,i} \qquad (22)$$

Taking the curl of $\boldsymbol{\beta}$, we obtain the compatibility condition for $\boldsymbol{\beta}$ as

$$\nabla \times \boldsymbol{\beta} = 0 \qquad \varepsilon_{ikl}\beta_{lj,k} = 0 \qquad (23)$$

This relation is the necessary condition for existence of $\mathbf{u}$, when $\boldsymbol{\beta}$ is known. In other words, the relation (23) is the requirement for integrability of $\boldsymbol{\beta}$ to obtain $\mathbf{u}$.

The infinitesimal strain tensor $\mathbf{e}$ and rotation tensor $\boldsymbol{\omega}$ are defined by decomposing the distortion tensor $\boldsymbol{\beta}$ to its symmetric and skew-symmetric parts as

$$\boldsymbol{\beta} = \mathbf{e} + \boldsymbol{\omega} \qquad \beta_{ij} = e_{ij} + \omega_{ij} \qquad (24)$$

where

$$\mathbf{e} = \boldsymbol{\beta}_{(\,)} = \frac{1}{2}\left(\boldsymbol{\beta} + \boldsymbol{\beta}^t\right) \qquad e_{ij} = \beta_{(ij)} = \frac{1}{2}\left(\beta_{ij} + \beta_{ji}\right) \qquad (25)$$

$$\mathbf{\omega} = \mathbf{\beta}_{[\,]} = \frac{1}{2}\left(\mathbf{\beta} - \mathbf{\beta}^t\right) \qquad \omega_{ij} = \beta_{[ij]} = \frac{1}{2}\left(\beta_{ij} - \beta_{ji}\right) \qquad (26)$$

respectively. Here we have introduced parentheses to denote the symmetric part of a second order tensor, whereas square brackets are associated with the skew-symmetric part.

We notice that these relations also can be written as

$$\mathbf{e} = \frac{1}{2}\left(\nabla \mathbf{u} + \mathbf{u}\nabla\right) \qquad e_{ij} = u_{(i,j)} = \frac{1}{2}\left(u_{j,i} + u_{i,j}\right) \qquad (27)$$

$$\mathbf{\omega} = \frac{1}{2}\left(\nabla \mathbf{u} - \mathbf{u}\nabla\right) \qquad \omega_{ij} = u_{[j,i]} = \frac{1}{2}\left(u_{j,i} - u_{i,j}\right) \qquad (28)$$

Since the true (polar) rotation tensor $\mathbf{\omega}$ is skew-symmetrical, one can introduce its corresponding dual pseudo (axial) rotation vector $\vec{\omega}$ as

$$\vec{\omega} = \frac{1}{2}\mathbf{\varepsilon} : \mathbf{\omega} = \frac{1}{2}\nabla \times \mathbf{u} \qquad \omega_i = \frac{1}{2}\varepsilon_{ijk}\omega_{jk} = \frac{1}{2}\varepsilon_{ijk}u_{k,j} \qquad (29)$$

Interestingly, there is also the twin relation

$$\mathbf{\omega} = \mathbf{\varepsilon} \bullet \vec{\omega} \qquad \omega_{ij} = \varepsilon_{ijk}\omega_k \qquad (30)$$

The definitions of $\mathbf{e}$ and $\vec{\omega}$ in (27) and (29) require

$$\nabla \times \mathbf{e} \times \nabla = 0 \qquad -\varepsilon_{ikl}\varepsilon_{jmn}e_{ln,km} = 0 \qquad (31)$$

$$\nabla \bullet \vec{\omega} = 0 \qquad \omega_{i,i} = 0 \qquad (32)$$

which are the compatibility conditions for $\mathbf{e}$ and $\vec{\omega}$, respectively. These conditions are the necessary conditions for existence of $\mathbf{u}$, when $\mathbf{e}$ or $\vec{\omega}$ are known. In general, the tensor $\nabla \times \mathbf{e} \times \nabla$ is called the Saint-Venant's or incompatibility tensor. As seen, this tensor vanishes for a compatible strain $\mathbf{e}$.

In continuum mechanics, we consider the rigid body portion of motion of infinitesimal elements of matter (or rigid triads) at each point of the continuum (Hadjesfandiari and Dargush, 2015). Therefore, the degrees of freedom are the displacements $\mathbf{u}$ and rotations $\vec{\omega}$ at each point, which describe translation and rotation of an infinitesimal element of matter in the neighborhood of the point, respectively. However, the continuity of matter within the continuum description restrains the rotation field $\vec{\omega}$ by the relation (29). This of course shows that the rotation field $\vec{\omega}$ is not independent of the displacement field $\mathbf{u}$.

By taking the curl of the skew-symmetric tensor $\boldsymbol{\omega}$, we obtain

$$\nabla \times \boldsymbol{\omega} = \nabla \times (\boldsymbol{\varepsilon} \bullet \vec{\omega}) \qquad \varepsilon_{ikl}\omega_{lj,k} = \varepsilon_{ikl}\varepsilon_{ljm}\omega_{m,k} \qquad (33)$$

or by using (21)

$$\nabla \times \boldsymbol{\omega} = (\nabla \bullet \vec{\omega})\mathbf{I} - \vec{\omega}\nabla \qquad \varepsilon_{ikl}\omega_{lj,k} = \omega_{k,k}\delta_{ij} - \omega_{i,j} \qquad (34)$$

Interestingly, by using the compatibility relation $\nabla \bullet \vec{\omega} = 0$, this compatiblity condition can also be written as

$$\nabla \times \boldsymbol{\omega} = -\vec{\omega}\nabla \qquad \varepsilon_{ikl}\omega_{lj,k} = -\omega_{i,j} \qquad (35)$$

The compatibility condition (23) for $\boldsymbol{\beta}$ can be written as

$$\nabla \times \mathbf{e} + \nabla \times \boldsymbol{\omega} = 0 \qquad \varepsilon_{ikl}e_{lj,k} + \varepsilon_{ikl}\omega_{lj,k} = 0 \qquad (36)$$

By using (34), the compatibility condition for $\boldsymbol{\beta}$ can be rewritten as

$$\nabla \times \mathbf{e} + (\nabla \bullet \vec{\omega})\mathbf{I} - \vec{\omega}\nabla = 0 \qquad \varepsilon_{ikl}e_{lj,k} + \omega_{k,k}\delta_{ij} - \omega_{i,j} = 0 \qquad (37)$$

while the transpose of this relation becomes

$$-\mathbf{e} \times \nabla + (\nabla \bullet \vec{\omega})\mathbf{I} - \nabla\vec{\omega} = 0 \qquad -\varepsilon_{jkl}e_{ki,l} + \omega_{k,k}\delta_{ij} - \omega_{j,i} = 0 \qquad (38)$$

By using the compatibility equation (32) for $\vec{\omega}$, i.e., $\nabla \bullet \vec{\omega} = 0$, the relations (37) and (38) reduce to

$$\nabla \times \mathbf{e} - \vec{\omega}\nabla = 0 \qquad \varepsilon_{ikl}e_{lj,k} - \omega_{i,j} = 0 \qquad (39)$$

and

$$-\mathbf{e} \times \nabla - \nabla\vec{\omega} = 0 \qquad -\varepsilon_{jkl}e_{ki,l} - \omega_{j,i} = 0 \qquad (40)$$

In small deformation theory, the bend-twist tensor is defined as

$$\mathbf{k} = \nabla\vec{\omega} \qquad k_{ij} = \omega_{j,i} \qquad (41)$$

We notice that

$$tr(\mathbf{k}) = \nabla \bullet \vec{\omega} = 0 \qquad k_{ii} = \omega_{i,i} = 0 \qquad (42)$$

which shows that the bend-twist tensor $\mathbf{k}$ is deviatoric.

By taking the curl of $\mathbf{k}$, we obtain the first compatibility condition for $\mathbf{k}$ as

$$\nabla \times \mathbf{k} = 0 \qquad \varepsilon_{ikl} k_{lj,k} = 0 \qquad (43)$$

This relation is the necessary condition for existence of $\vec{\omega}$, when $\mathbf{k}$ is given.

We notice that the compatibility condition (36) for distortion $\boldsymbol{\beta}$ can also be written as

$$\nabla \times \mathbf{e} + \boldsymbol{\varepsilon} : (\mathbf{k} \bullet \boldsymbol{\varepsilon}) = 0 \qquad \varepsilon_{ikl} e_{lj,k} + \varepsilon_{ikl} \varepsilon_{ljm} k_{km} = 0 \qquad (44)$$

or

$$\nabla \times \mathbf{e} + tr(\mathbf{k})\mathbf{I} - \mathbf{k}^t = 0 \qquad \varepsilon_{ikl} e_{lj,k} + k_{ll} \delta_{ij} - k_{ji} = 0 \qquad (45)$$

Interestingly, this important relation necessitates the deviatoric character of the bend-twist tensor, i.e. $tr(\mathbf{k}) = 0$, by itself, and demonstrates the interrelationship between the bend-twist tensor $\mathbf{k}$ and the strain tensor $\mathbf{e}$.

The second compatibility condition for $\mathbf{k}$ is obtained by transposing the relation (45) as

$$\mathbf{k} - tr(\mathbf{k})\mathbf{I} = -\mathbf{e} \times \nabla \qquad k_{ij} - k_{ll} \delta_{ij} = \varepsilon_{jkl} e_{li,k} \qquad (46)$$

By using the deviatoric compatibility condition (42), i.e., $tr(\mathbf{k}) = 0$, for the bend-twist tensor $\mathbf{k}$, this relation can also be written as

$$\mathbf{k} = -\mathbf{e} \times \nabla \qquad k_{ij} = \varepsilon_{jkl} e_{il,k} \qquad (47)$$

This compatibility condition represents the interrelationship between $\mathbf{e}$ and $\mathbf{k}$, indicating that the bend-twist tensor $\mathbf{k}$ is not independent of the strain $\mathbf{e}$. Notice that by using this relation in the first compatibility (43) for $\mathbf{k}$, we obtain the strain compatibility relation (31) for $\mathbf{e}$.

As its name indicates, the bend-twist tensor $\mathbf{k}$ describes the combination of bending and twisting of elements of the continuum. deWit (1970) has stated that the diagonal components of this tensor describe a twisting and the off-diagonal ones a bending deformations. However, these two sets of diagonal and off-diagonal components do not produce consistent torsion and curvature tensors. In particular, the components depend upon the orientation of the reference axes. Consequently, by taking diagonal components or deleting these components from the bend-twist tensor $\mathbf{k}$, one cannot create valid tensors. On the other hand, the symmetric and skew-symmetric parts of the bend-twist tensor $\mathbf{k}$ produce valid tensors with twisting and bending characters, respectively (Hadjesfandiari and Dargush, 2011; Hadjesfandiari at al., 2015). By decomposing the bend-twist tensor $\mathbf{k}$ into symmetric and skew-symmetric parts, we obtain

$$\mathbf{k} = \boldsymbol{\chi} + \boldsymbol{\kappa} \qquad k_{ij} = \chi_{ij} + \kappa_{ij} \qquad (48)$$

where the symmetric torsion tensor $\boldsymbol{\chi}$ and the skew-symmetric mean curvature tensor $\boldsymbol{\kappa}$ (Hadjesfandiari and Dargush, 2011, 2015) are

$$\boldsymbol{\chi} = \mathbf{k}_{(\ )} = \frac{1}{2}\left(\mathbf{k} + \mathbf{k}^t\right) \qquad \chi_{ij} = k_{(ij)} = \frac{1}{2}\left(k_{ij} + k_{ji}\right) \qquad (49)$$

$$\boldsymbol{\kappa} = \mathbf{k}_{[\ ]} = \frac{1}{2}\left(\mathbf{k} - \mathbf{k}^t\right) \qquad \kappa_{ij} = k_{[ij]} = \frac{1}{2}\left(k_{ij} - k_{ji}\right) \qquad (50)$$

It should be noticed that the aforementioned statement of deWit is valid only for the representation of the bend-twist tensor $\mathbf{k}$ in the principal directions of the symmetric torsion tensor $\boldsymbol{\chi}$.

Interestingly, the tensors $\boldsymbol{\chi}$ and $\boldsymbol{\kappa}$ can be written as

$$\boldsymbol{\chi} = \frac{1}{2}\left(\nabla \vec{\omega} + \vec{\omega} \nabla\right) \qquad \chi_{ij} = \omega_{(j,i)} = \frac{1}{2}\left(\omega_{j,i} + \omega_{i,j}\right) \qquad (51)$$

$$\boldsymbol{\kappa} = \frac{1}{2}\left(\nabla \vec{\omega} - \vec{\omega} \nabla\right) \qquad \kappa_{ij} = \omega_{[j,i]} = \frac{1}{2}\left(\omega_{j,i} - \omega_{i,j}\right) \qquad (52)$$

Since the pseudo (axial) mean curvature tensor $\boldsymbol{\kappa}$ is skew-symmetrical, one also can introduce its corresponding dual true (polar) mean curvature vector $\vec{\kappa}$ as

$$\vec{\kappa} = \frac{1}{2}\boldsymbol{\varepsilon} : \boldsymbol{\kappa} \qquad \kappa_i = \frac{1}{2}\varepsilon_{ijk}\kappa_{jk} \qquad (53)$$

with the dual relation

$$\boldsymbol{\kappa} = \boldsymbol{\varepsilon} \bullet \vec{\kappa} \qquad \kappa_{ij} = \varepsilon_{ijk}\kappa_k \qquad (54)$$

The mean curvature vector also can be defined as

$$\vec{\kappa} = \frac{1}{2}\nabla \times \vec{\omega} \qquad \kappa_i = \frac{1}{2}\varepsilon_{ijk}\omega_{k,j} \qquad (55)$$

which clearly shows

$$\nabla \bullet \vec{\kappa} = 0 \qquad \kappa_{i,i} = 0 \qquad (56)$$

We notice that this is the compatibility condition for $\vec{\kappa}$. This relation is the necessary condition for existence of $\vec{\omega}$, when $\vec{\kappa}$ is known.

It is also found that

$$\vec{\kappa} = -\frac{1}{2}\nabla \bullet \boldsymbol{\omega} = \frac{1}{4}\left(\nabla\nabla \bullet \mathbf{u} - \nabla^2 \mathbf{u}\right) \qquad \kappa_i = -\frac{1}{2}\omega_{ji,j} = \frac{1}{4}\left(u_{j,ji} - \nabla^2 u_i\right) \qquad (57)$$

By using the second compatibility relation (46) for the bend-twist tensor $\mathbf{k}$, we obtain

$$\mathbf{k} = tr(\mathbf{k})\mathbf{I} - \mathbf{e} \times \nabla \qquad k_{ij} = k_{ll}\delta_{ij} + \varepsilon_{jkl}e_{il,k} \qquad (58)$$

Consequently, by using this relation, we can express the torsion tensor $\boldsymbol{\chi}$ and mean curvature tensor $\boldsymbol{\kappa}$ in terms of the strain tensor $\mathbf{e}$ as

$$\boldsymbol{\chi} = tr(\mathbf{k})\mathbf{I} + \frac{1}{2}(-\mathbf{e}\times\nabla + \nabla\times\mathbf{e}) \qquad \chi_{ij} = k_{ll}\delta_{ij} + \frac{1}{2}\left(\varepsilon_{jkl}e_{il,k} + \varepsilon_{ikl}e_{jl,k}\right) \qquad (59)$$

$$\boldsymbol{\kappa} = -\frac{1}{2}(\mathbf{e}\times\nabla + \nabla\times\mathbf{e}) \qquad \kappa_{ij} = \frac{1}{2}\left(\varepsilon_{jkl}e_{il,k} - \varepsilon_{ikl}e_{jl,k}\right) \qquad (60)$$

We notice that the relation (59) results in

$$tr(\boldsymbol{\chi}) = 3\,tr(\mathbf{k}) \qquad \chi_{ii} = 3k_{ll} \qquad (61)$$

For a compatible deformation, the deviatoric compatibility condition (42) of the bend-twist tensor, i.e., $tr(\mathbf{k}) = 0$, requires the deviatoric compatibility character of the torsion tensor

$$tr(\boldsymbol{\chi}) = 0 \qquad \chi_{ii} = 0 \qquad (62)$$

Therefore, by using this condition, the relation (59) reduces to

$$\chi = \frac{1}{2}\left(-\mathbf{e}\times\nabla + \nabla\times\mathbf{e}\right) \qquad \chi_{ij} = \frac{1}{2}\left(\varepsilon_{jkl}e_{il,k} + \varepsilon_{ikl}e_{jl,k}\right) \tag{63}$$

which demonstrates the interrelationship between the compatible torsion tensor $\chi$ and strain tensor $\mathbf{e}$.

Interestingly, the mean curvature vector $\vec{\kappa}$ can also be expressed in terms of the strain tensor $\mathbf{e}$ as

$$\vec{\kappa} = \frac{1}{2}\left[\nabla tr(\mathbf{e}) - \nabla\bullet\mathbf{e}\right] \qquad \kappa_i = \frac{1}{2}\left(e_{kk,i} - e_{ki,k}\right) \tag{64}$$

The mean curvature tensor $\kappa$ and the torsion tensor $\chi$ are of fundamental importance in consistent couple stress theory, which will be presented in Section 5. The relations (60) and (63) between these tensors and the strain tensor $\mathbf{e}$ play a critical role in developing the new C-CDT, as will be seen in Section 8.

## 4. Discrete defect theory

### 4.1. Compatibility conditions and integrability

Let us assume the strain tensor $\mathbf{e}$ and the bend-twist tensor $\mathbf{k}$ are known for a deformed body. For the existence of fields of displacement $\mathbf{u}$ and rotation $\vec{\omega}$, it is required that the deformation be compatible. This means $\mathbf{e}$ and $\mathbf{k}$ must satisfy the compatibility conditions (43) and (45), which are presented again as

$$\nabla\times\mathbf{k} = 0 \qquad \varepsilon_{ikl}k_{lj,k} = 0 \tag{65}$$

$$\nabla\times\mathbf{e} + tr(\mathbf{k})\mathbf{I} - \mathbf{k}^t = 0 \qquad \varepsilon_{ikl}e_{lj,k} + k_{ll}\delta_{ij} - k_{ji} = 0 \tag{66}$$

We should note that these conditions already include compatibility conditions (31) and (42). As demonstrated, the compatibility condition (66) also can be written in the form (44). It should be emphasized that the compatibility conditions (65) and (66) are complete and do not need the inclusion of other compatibility conditions, such as (31) and (42). This form of the compatibility conditions is very important in our developments in this paper, regarding Weingarten's theorem and the definition of internal continuous defect tensors in continuous defect theory.

The compatibility conditions (65) and (66) are also sufficient when the body is simply connected (deWit, 1970). However, these conditions are not sufficient for multiply connected bodies, where the displacement **u** and rotation $\vec{\omega}$ fields may be multiple-valued. Recall that a body is simply connected when all simple closed curves or circuits in its region can be continuously shrunk or reduced to a point without going outside the body. Otherwise, when any simple curve is not reducible, the body is multiply connected.

Weingarten (1901) and deWit (1970) have derived explicit expressions for the displacement **u** and rotation $\vec{\omega}$ when the compatible fields **k** and **e** are given. It turns out that it is only necessary to know $\vec{\omega}_0$ and $\mathbf{u}_0$ at some arbitrary point $\mathbf{r}_0$ to find $\vec{\omega}$ and **u** at any arbitrary point **r**. For $\vec{\omega}$ we have

$$\vec{\omega}(\mathbf{r}) = \vec{\omega}(\mathbf{r}_0) + \int_{\mathbf{r}_0}^{\mathbf{r}} d\vec{\omega} \qquad \omega_i = \omega_{0i} + \int_{x_{i0}}^{x_i} d\omega_i' \qquad (67)$$

or

$$\vec{\omega}(\mathbf{r}) = \vec{\omega}_0 + \int_{\mathbf{r}_0}^{\mathbf{r}} d\mathbf{r}' \bullet \mathbf{k}(\mathbf{r}') \qquad \omega_i = \omega_{0i} + \int_{x_{i0}}^{x_i} k_{ji}' dx_j' \qquad (68)$$

deWit (1970) has also derived the expression for **u** as

$$\begin{aligned}
\mathbf{u}(\mathbf{r}) &= \mathbf{u}_0 + \int_{\mathbf{r}_0}^{\mathbf{r}} d\mathbf{r}' \bullet \boldsymbol{\beta}(\mathbf{r}') \\
&= \mathbf{u}_0 + \int_{\mathbf{r}_0}^{\mathbf{r}} d\mathbf{r}' \bullet [\mathbf{e}' + \boldsymbol{\omega}'] \\
&= \mathbf{u}_0 + \int_{\mathbf{r}_0}^{\mathbf{r}} [d\mathbf{r}' \bullet \mathbf{e}' + \vec{\omega}' \times d\mathbf{r}'] \\
&= \mathbf{u}_0 + \int_{\mathbf{r}_0}^{\mathbf{r}} [d\mathbf{r}' \bullet \mathbf{e}' - \vec{\omega}' \times d(\mathbf{r} - \mathbf{r}')] \\
&= \mathbf{u}_0 + \int_{\mathbf{r}_0}^{\mathbf{r}} \{d\mathbf{r}' \bullet \mathbf{e}' - d[\vec{\omega}' \times d(\mathbf{r} - \mathbf{r}')] + d\vec{\omega}' \times (\mathbf{r} - \mathbf{r}')\}
\end{aligned} \qquad (69)$$

$$u_i = u_{0i} + \int_{x_{i0}}^{x_i} \beta'_{ji} dx'_j$$

$$= u_{0i} + \int_{x_{i0}}^{x_i} \left( e'_{ji} + \omega'_{ji} \right) dx'_j$$

$$= u_{0i} + \int_{x_{i0}}^{x_i} \left( e'_{ji} + \varepsilon_{jik} \omega'_k \right) dx'_j$$

$$= u_{0i} + \int_{x_{i0}}^{x_i} \left[ e'_{ji} dx'_j - \varepsilon_{ikj} \omega'_k d\left(x_j - x'_j\right) \right]$$

$$= u_{0i} + \int_{x_{i0}}^{x_i} \left\{ e'_{ji} dx'_j - d\left[ \varepsilon_{ikj} \omega'_k \left(x_j - x'_j\right) \right] + \varepsilon_{ikj} d\omega'_k \left(x_j - x'_j\right) \right\}$$

or

$$\mathbf{u}(\mathbf{r}) = \mathbf{u}_0 + \vec{\omega}_0 \times (\mathbf{r} - \mathbf{r}_0) + \int_{\mathbf{r}_0}^{\mathbf{r}} \left\{ d\mathbf{r}' \bullet \mathbf{e}' + (d\mathbf{r}' \bullet \mathbf{k}') \times (\mathbf{r} - \mathbf{r}') \right\} \tag{70}$$

$$u_i = u_{0i} + \varepsilon_{ikj} \omega_{0k} \left(x_j - x_{0j}\right) + \int_{x_{i0}}^{x_i} \left[ e'_{li} + \varepsilon_{ikj} k'_{lk} \left(x_j - x'_j\right) \right] dx'_l$$

These results for the rotation $\vec{\omega}$ and displacement $\mathbf{u}$ are single valued if the line integrals in (68) and (70) are independent of the path integral around a closed circuit $C$ starting and ending at point $\mathbf{r}_0$. For this it is required that these integrals vanish for every arbitrary closed curve or circuit $C$ in the body, that is

$$\oint_C d\mathbf{r}' \bullet \mathbf{k}'(\mathbf{r}') = 0 \qquad \oint_C k'_{ji} dx'_j = 0 \tag{71}$$

$$\oint_C d\mathbf{r}' \bullet \left[ \mathbf{e}' + \mathbf{k}' \times (\mathbf{r} - \mathbf{r}') \right] = 0 \qquad \oint_C \left[ e'_{li} + \varepsilon_{ikj} k'_{lk} \left(x_j - x'_j\right) \right] dx'_l = 0 \tag{72}$$

If the body is simply connected, then we can use Stokes' theorem to transform these circuit integrals to

$$\oint_C d\mathbf{r}' \bullet \mathbf{k}(\mathbf{r}') = \int_S d\mathbf{S}' \bullet \nabla' \times \mathbf{k}' \qquad \oint_C k'_{ji} dx'_j = \int_S \varepsilon_{jkl} k'_{li,k} n'_j dS' \tag{73}$$

$$\oint_C d\mathbf{r}' \bullet \left[ \mathbf{e}' + \mathbf{k}' \times (\mathbf{r} - \mathbf{r}') \right] = \int_S d\mathbf{S}' \bullet \nabla' \times \left[ \mathbf{e}' + \mathbf{k}' \times (\mathbf{r} - \mathbf{r}') \right]$$

$$= \int_S d\mathbf{S}' \bullet \left[ \nabla' \times \mathbf{e}' + tr(\mathbf{k}') \mathbf{I} - \mathbf{k}'^t + (\nabla' \times \mathbf{k}') \times (\mathbf{r} - \mathbf{r}') \right] \tag{74}$$

$$\oint_C \left[ e'_{li} + \varepsilon_{ikj} k'_{lk} \left( x_j - x'_j \right) \right] dx'_l = \int_S \varepsilon_{pql} \left[ e'_{li} + \varepsilon_{ikj} k'_{lk} \left( x_j - x'_j \right) \right]_{,q} n'_p dS'$$

$$= \int_S \left[ \varepsilon_{pql} e'_{li,q} + k'_{ll} \delta_{pi} - k'_{ip} + \varepsilon_{pql} \varepsilon_{ikj} k'_{lk,q} \left( x_j - x'_j \right) \right] n'_p dS'$$

where $S$ is any arbitrary oriented surface in the body bounded by the circuit $C$ with positive direction. It should be noticed that in these relations

$$d\mathbf{S} = dS\mathbf{n} \qquad dS_i = dSn_i \qquad (75)$$

where $dS$ is the surface element with unit outward normal vector $\mathbf{n}$. Therefore, for a simply connected body, $\vec{\omega}$ and $\mathbf{u}$ are single valued if and only if

$$\oint_C d\mathbf{r}' \bullet \mathbf{k}(\mathbf{r}') = \int_S d\mathbf{S}' \bullet \nabla' \times \mathbf{k}' = 0 \qquad (76)$$

$$\oint_C d\mathbf{r}' \bullet \left[ \mathbf{e}' + \mathbf{k}' \times (\mathbf{r} - \mathbf{r}') \right]$$
$$= \int_S d\mathbf{S}' \bullet \left[ \nabla' \times \mathbf{e}' + tr(\mathbf{k}')\mathbf{I} - \mathbf{k}'^t \right] + \int_S d\mathbf{S}' \bullet \left[ (\nabla' \times \mathbf{k}') \times (\mathbf{r} - \mathbf{r}') \right] = 0 \qquad (77)$$

This result shows that for a simply connected body the compatibility conditions (65) and (66) are necessary and sufficient conditions for integrability of $\mathbf{k}$ and $\mathbf{e}$ to obtain the rotations $\vec{\omega}$ and displacements $\mathbf{u}$. This means for a simply connected body the rotation vector $\vec{\omega}$ and displacement vector $\mathbf{u}$ are single-valued functions.

For a multiply connected body the compatibility conditions (65) and (66) are not sufficient to establish the single-valued character of the rotation vector $\vec{\omega}$ and displacement vector $\mathbf{u}$, if the circuit $C$ is irreducible. For an irreducible circuit $C$, $\vec{\omega}$ and $\mathbf{u}$ may not return to their original values on following the circuit around. The changes are given by setting $\mathbf{r} = \mathbf{r}_0$ in the general expressions (68) and (70) as

$$[\vec{\omega}] = \oint_C d\mathbf{r}' \bullet \mathbf{k}' \qquad [\omega_i] = \oint_C k'_{ji} dx'_j \qquad (78)$$

$$[\mathbf{u}] = \oint_C d\mathbf{r}' \bullet \left[ \mathbf{e}' - \mathbf{k}' \times \mathbf{r}' \right] + \left( \oint_C d\mathbf{r}' \bullet \mathbf{k}' \right) \times \mathbf{r} \qquad [u_i] = \oint_C \left[ e'_{li} - \varepsilon_{ikj} k'_{lk} x'_j \right] dx'_l + \varepsilon_{ikj} x_j \oint_C k'_{lk} dx'_l \qquad (79)$$

The jumps $[\vec{\omega}]$ and $[\mathbf{u}]$ represent the discontinuity of the rotation vector $\vec{\omega}$ and the displacement vector at $\mathbf{r} = \mathbf{r}_0$, respectively. We notice that these expressions can be written as

$$[\vec{\omega}] = \vec{\Omega}_O \qquad\qquad [\omega_i] = \Omega_{Oi} \qquad (80)$$

$$[\mathbf{u}] = \mathbf{B}_O + \vec{\Omega}_O \times \mathbf{r} \qquad [u_i] = B_{Oi} + \varepsilon_{ikj}\Omega_{Ok}x_j \qquad (81)$$

where we have defined the Frank true (polar) vector $\vec{\Omega}_O$ and Burgers pseudo (axial) vector $\mathbf{B}_O$ relative to the origin as

$$\vec{\Omega}_O = \oint_C d\mathbf{r}' \bullet \mathbf{k}' \qquad\qquad \Omega_{Oi} = \oint_C k'_{ji} dx'_j \qquad (82)$$

$$\mathbf{B}_0 = \oint_C d\mathbf{r}' \bullet [\mathbf{e}' - \mathbf{k}' \times \mathbf{r}'] \qquad B_{Oi} = \oint_C \left[ e'_{li} - \varepsilon_{ikj} k'_{lk} x'_j \right] dx'_l \qquad (83)$$

It should be emphasized that the rotation vector $\vec{\omega}$ is a pseudo (axial) vector, whereas the displacement vector $\mathbf{u}$ is a true (polar) vector. The true character of the Frank vector $\vec{\Omega}_O$ and pseudo character of the Burgers vector $\mathbf{B}_O$ are the result of the fact that the jumps $[\vec{\omega}]$ and $[\mathbf{u}]$ depend on the positive or negative orientation of the circuit $C$ in the line integrals (78) and (79). It should be noticed that the Frank vector $\vec{\Omega}_O$ and Burgers vector $\mathbf{B}_0$ remain invariant even if the circuit $C$ is replaced with any other reconcilable curve $C_0$, which means it can be deformed continuously into $C$, without leaving the body (deWit, 1970). This is the result of the compatibility conditions (65) and (66) in Stokes' theorem for a multiply connected body.

### 4.2. Weingarten's theorem and its consequences

For a simply connected body every circuit is reducible, which means the circuit $C$ can be deformed into a point. As a result, we have $[\vec{\omega}] = \vec{\Omega}_O = 0$, and $[\mathbf{u}] = \mathbf{B}_O = 0$, which agrees with the results in (71) and (72) for single valuedness of $\vec{\omega}$ and $\mathbf{u}$. However, for a multiply connected body, when the circuit $C$ encircles a hole, we notice that the circuit $C$ is irreducible. Therefore, the Frank vector $\vec{\Omega}_O$ and Burgers vector $\mathbf{B}_O$ do not necessarily vanish, but are constant for all reconcilable circuits. For this case the vectors $\vec{\omega}$ and $\mathbf{u}$ might be multiple-valued.

Interestingly, the relations (80) and (81) express Weingarten's theorem (Weingarten, 1901; Love, 1920) which states:

On following around an irreducible circuit in a multiply connected body with a deformation satisfying the compatibility conditions (65) and (66), the rotation vector $\vec{\omega}$ and displacement vector $\mathbf{u}$ change by an amount that would be possible for a rigid body.

If the origin is in the region of the body, we notice that the jumps $[\vec{\omega}]$ and $[\mathbf{u}]$ at the origin are

$$[\vec{\omega}](\mathbf{r}=0) = \vec{\Omega}_O \tag{84}$$

$$[\mathbf{u}](\mathbf{r}=0) = \mathbf{B}_O \tag{85}$$

However, based on Weingarten's theorem, the jumps $[\vec{\omega}]$ and $[\mathbf{u}]$ at an arbitrary point specified with $\mathbf{r}$ are

$$[\vec{\omega}](\mathbf{r}) = \vec{\Omega} = \vec{\Omega}_0 \tag{86}$$

$$[\mathbf{u}](\mathbf{r}) = \mathbf{B} = \mathbf{B}_0 + \vec{\Omega} \times \mathbf{r} \tag{87}$$

Therefore, Weingarten's theorem allows us to represent the discrete defect or rigid body jump by using a system of Frank and Burgers vectors at any point. We notice the effect of a system consisting of a Frank true vector $\vec{\Omega}_O$ and a Burgers pseudo vector $\mathbf{B}_O$ at the origin can be replaced by the equivalent system of Frank true vector $\vec{\Omega} = \vec{\Omega}_O$ and Burgers pseudo vector $\mathbf{B} = \mathbf{B}_0 + \vec{\Omega} \times \mathbf{r}$ at point $\mathbf{r}$. It should be emphasized that the effect of these systems of Frank-Burgers vectors are equivalent in the sense that they create the same discontinuities or jumps $[\vec{\omega}]$ and $[\mathbf{u}]$.

Interestingly, this character of the system of Frank-Burgers vectors is analogous to the character of a force-couple system in rigid body mechanics. The system consisting of a force vector $\mathbf{F}_O$ and a couple force with moment $\mathbf{M}_O$ (free moment $\mathbf{M}_O$) at the origin can be replaced by an equipollent force-couple system $\mathbf{F}$ and $\mathbf{M}$ at point $\mathbf{r}$, where

$$\mathbf{F} = \mathbf{F}_O \tag{88}$$

$$\begin{aligned}\mathbf{M} &= \mathbf{M}_0 + (-\mathbf{r}) \times \mathbf{F}_O \\ &= \mathbf{M}_0 + \mathbf{F}_O \times \mathbf{r}\end{aligned} \tag{89}$$

As known, the dynamic or static behavior of the rigid body is the same under any of these equivalent or equipollent force-couple systems. However, we notice that they create different internal stresses and deformations in the body.

Therefore, Weingarten's theorem shows that the overall effect of Frank true vector $\vec{\Omega}$ and Burgers pseudo vector $\mathbf{B}$ at any point is analogous to the effect of force true-vector $\mathbf{F}$ and free moment vector $\mathbf{M}$.

For the general case, where the defect is specified by the system of a Frank true vector $\vec{\Omega}_O$ and Burgers pseudo vector $\mathbf{B}_0$ at the origin, we have the following results:

1. The kinematical effect of Burgers vector $\mathbf{B}_O$ is independent of its position, and thus can be considered as a free vector.
2. Although the effect of a Frank vector $\vec{\Omega}_O$ does not change by moving it along its line of action, it brings an extra Burgers vector $(-\mathbf{r}) \times \vec{\Omega}_O = \vec{\Omega}_O \times \mathbf{r}$ to the new point at $\mathbf{r}$, where

$$\vec{\Omega} = \vec{\Omega}_O \tag{90}$$

$$\begin{aligned}\mathbf{B} &= \mathbf{B}_O + \vec{\Omega}_O \times \mathbf{r} \\ &= \mathbf{B}_O + (-\mathbf{r}) \times \vec{\Omega}_O\end{aligned} \tag{91}$$

Now, we consider the following two fundamental special cases:

*Case 1*. There is no Burgers vector at the origin, that is, $\mathbf{B}_O = 0$. Consequently, the jumps $[\vec{\omega}]$ and $[\mathbf{u}]$ at arbitrary point $\mathbf{r}$ are

$$[\vec{\omega}](\mathbf{r}) = \vec{\Omega} = \vec{\Omega}_0 \tag{92}$$

$$[\mathbf{u}](\mathbf{r}) = \mathbf{B} = \vec{\Omega} \times \mathbf{r} \tag{93}$$

Weingarten's theorem shows that the displacement vector $\mathbf{u}$ changes by an amount that would be possible for a rigid body rotation around the origin.

*Case 2*. There is no Frank vector at the origin, that is $\vec{\Omega}_O = 0$. Subsequently, the jumps $[\vec{\omega}]$ and $[\mathbf{u}]$ at $\mathbf{r}$ are

$$[\vec{\omega}](\mathbf{r}) = \vec{\Omega} = 0 \tag{94}$$

$$[\mathbf{u}](\mathbf{r}) = \mathbf{B} = \mathbf{B}_O \tag{95}$$

Therefore, for this case, there is no jump in the rotation vector, $[\vec{\omega}] = 0$, but there is a constant jump in the displacement, $[\mathbf{u}] = \mathbf{B} = \mathbf{B}_O$, for all points. Based on Weingarten's theorem, the displacement vector $\mathbf{u}$ changes by an amount that would be possible for a rigid body translation. Therefore, the vector $\mathbf{B}_O$ can be considered as a free vector in this case.

Remarkably, we notice that the effect of the given free Burgers pseudo vector $\mathbf{B}_O$ is equivalent to the effect of a couple of arbitrary Frank true vectors $\vec{\Omega}^I$ and $\vec{\Omega}^{II} = -\vec{\Omega}^I$ at positions $\mathbf{r}^I$ and $\mathbf{r}^{II} = \mathbf{r}^I + \mathbf{d}$, respectively, as long as

$$\begin{aligned}\mathbf{B} = \mathbf{B}_O &= \vec{\Omega}^{II} \times \mathbf{d} \\ &= \mathbf{d} \times \vec{\Omega}^I\end{aligned} \tag{96}$$

Here $\mathbf{d}$ represents the relative position of vector $\vec{\Omega}^{II}$ with respect to $\vec{\Omega}^I$. The overall effect of couple vectors $\vec{\Omega}^I$ and $\vec{\Omega}^{II} = -\vec{\Omega}^I$ is independent of their values and positions, but depends on their resultant Burgers vector $\mathbf{B}_O = \mathbf{d} \times \vec{\Omega}^I$. This allows us to represent the Burgers couple vectors system by its resultant Burgers vector $\mathbf{B} = \mathbf{B}_O = \mathbf{d} \times \vec{\Omega}^I$ as a free vector, which will be denoted by the symbol $\mathbf{b}$ for emphasis in the following. However, we notice that these equivalent systems may create different deformations (strains and curvatures) and internal stresses in the body. For simplicity, in the following, we use the term Burgers vector, instead of Burgers couple vector.

### 4.3. Discrete disclinations and dislocations

Based on Weingarten's theorem, the components of Frank vector $\vec{\Omega}$ and Burgers vector $\mathbf{B} = \mathbf{b}$ represent six types of discrete line defects in the body. These six types of geometrically necessary crystal line defects introduced by Volterra (1907), include three rotational jumps or disclinations $\vec{\Omega}_1$, $\vec{\Omega}_2$ and $\vec{\Omega}_3$ around the $x_1 x_2 x_3$ coordinate axes, respectively, and three pure translational jumps or dislocations $\mathbf{B}_1 = \mathbf{b}_1$, $\mathbf{B}_2 = \mathbf{b}_2$ and $\mathbf{B}_3 = \mathbf{b}_3$ in the direction of the coordinate axes. Interestingly, Love (1920) and Frank (1958) introduced the terminology of dislocation $\mathbf{B} = \mathbf{b}$ and disclination $\vec{\Omega}$, respectively, in order to specify the states created by the two kinds of defects.

Volterra (1907) discussed the characteristics of these defects in detail by cutting a hollow circular cylinder. Let the axis of the cylinder be along the $x_3$ direction, while the cut lies in the $x_1 x_3$ plane, as shown in Figure 1.

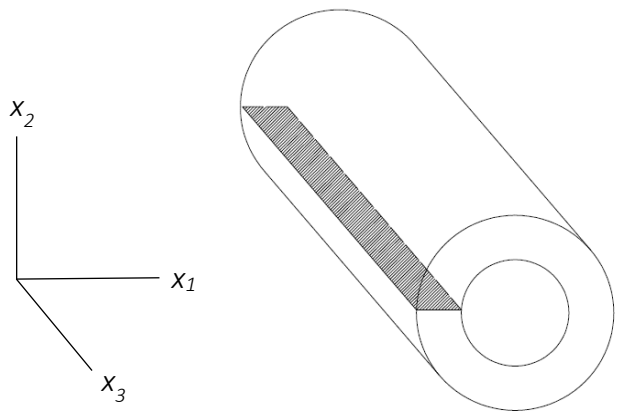

**Figure 1.** The original cut cylinder and coordinate system.

Figures 2 and 3 represent the three kinds of discrete disclinations and three kinds of discrete dislocations, respectively. Note that we have deliberately presented the disclinations before dislocations. The reason for this choice will be clarified later in the paper.

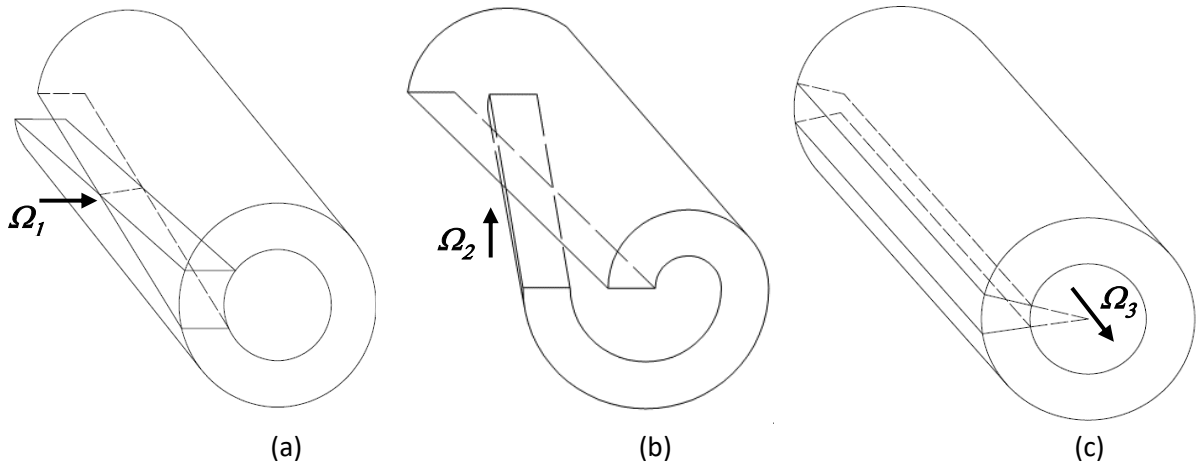

**Figure 2.** Three discrete disclinations: (a) tilting edge disclination, (b) twisting edge disclination, (c) wedge disclination.

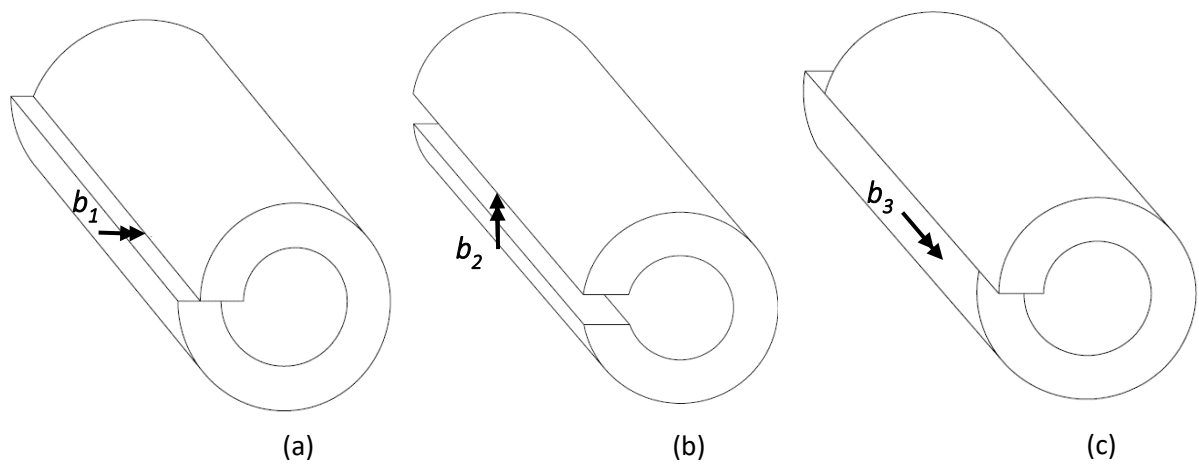

**Figure 3.** Three discrete dislocations: (a) gliding edge dislocation, (b) climbing edge dislocation, (c) screw dislocation.

The three types of discrete disclinations consist of rotating or inclining of the two sides of the cut relative to each other by vectors $\vec{\Omega}_1$, $\vec{\Omega}_2$ and $\vec{\Omega}_3$, sealing them together and removing, if present, the forces and moments that have brought the sides to the sealing position. The discrete disclinations in Figure 2a,b, whose Frank vectors $\vec{\Omega}_1$ and $\vec{\Omega}_2$ are perpendicular to the defect line, are identified as edge disclinations, where $\vec{\Omega}_1$ is a tilting edge disclination, and $\vec{\Omega}_2$ is a twisting edge disclination. The disclination in Figure 2c, whose Frank vector $\vec{\Omega}_3$ is parallel to the defect line, is a wedge disclination.

We also notice that the three types of discrete dislocations consist of displacing the two sides of the cut relative to each other by vectors $\mathbf{b}_1$, $\mathbf{b}_2$ and $\mathbf{b}_3$, sealing them together and removing, if present, the forces and moments that have brought the sides to the sealing position. The discrete dislocations in Figure 3a,b whose Burger vectors $\mathbf{b}_1$ and $\mathbf{b}_2$ are perpendicular to the defect line are identified as edge dislocations, where $\mathbf{b}_1$ is a gliding edge dislocation, and $\mathbf{b}_2$ is a climbing dislocation. The dislocation in Figure 3c whose Burger vector $\mathbf{b}_3$ is parallel to the defect line is identified as a screw dislocation.

In the gliding edge dislocation $\mathbf{b}_1$ and screw dislocation $\mathbf{b}_3$, the two planes on the sides of the cut contain both the dislocation line and Burgers vector. Therefore, these two types of dislocations are the result of slipping of planes cut on each other. However, the mechanism of movement is fundamentally different for the climbing edge dislocation, where the Burgers vector $\mathbf{b}_2$ is perpendicular to the dislocation line and two planes on the sides of the cut. Therefore, this type of dislocation allows an edge to move perpendicular to its cut planes.

In this section, the Frank vector $\vec{\Omega}$ and Burgers vector $\mathbf{B}$ (free Burgers vector) have been introduced as line defects, which are classified as discrete disclinations and dislocations, respectively, in multiply connected bodies. As demonstrated, these vectors are analogous to the force and couple (free moment) in mechanics of rigid bodies, respectively. This result suggest that we can consider the concentrated Frank vector $\vec{\Omega}$ and Burgers vector $\mathbf{B}$ at any point in the body analogous to the concentrated force and couple in mechanics, respectively. The mechanism behind the physical concentrated Frank vector $\vec{\Omega}$ and Burgers vector $\mathbf{B}$ will be investigated in Section 6. It will be seen that these concentrated external vectors can create plastic deformation in the body, which in turn results in internal continuous disclination and dislocation distributions in the body. Therefore, discrete defect theory motivates the development of continuous defect theory with internal continuous disclination and dislocation density tensors. However, we first consider the state of stresses in the body in the next section by presenting a review of couple stress theory, which plays a crucial role in developing consistent continuous defect theory (C-CDT) in Section 8.

## 5. Couple stress theory

### 5.1. General formulation

Consider the material continuum occupying the volume $V_0$ bounded by the boundary surface $S_0$ under the influence of external surface loading that produces internal stresses in the body. In the more general size-dependent continuum mechanics theory, one goes beyond the classical Cauchy hypothesis and assumes that the interaction acting on a surface element $dS$ with unit normal vector $\mathbf{n}$ is specified by means of a force vector $d\mathbf{F} = \mathbf{t}^{(n)} dS$ and a couple vector $d\mathbf{M} = \mathbf{m}^{(n)} dS$, where $\mathbf{t}^{(n)}$ is the force-traction true (polar) vector and $\mathbf{m}^{(n)}$ is the couple-traction pseudo (axial) vector. Therefore, the internal stresses are represented by generally non-symmetric true (polar) force-stress $\boldsymbol{\sigma}$ and pseudo (axial) couple-stress $\boldsymbol{\mu}$ tensors (Cosserat and Cosserat, 1909), where

$$\mathbf{t}^{(n)} = \mathbf{n} \bullet \boldsymbol{\sigma} \qquad t_i^{(n)} = \sigma_{ji} n_j \tag{97}$$

$$\mathbf{m}^{(n)} = \mathbf{n} \bullet \boldsymbol{\mu} \qquad m_i^{(n)} = \mu_{ji} n_j \tag{98}$$

The force- and couple- stress tensors in the original theory are non-symmetric and can be generally decomposed into symmetric and skew-symmetric parts

$$\boldsymbol{\sigma} = \boldsymbol{\sigma}_{(\,)} + \boldsymbol{\sigma}_{[\,]} \qquad \sigma_{ij} = \sigma_{(ij)} + \sigma_{[ij]} \tag{99}$$

$$\boldsymbol{\mu} = \boldsymbol{\mu}_{(\,)} + \boldsymbol{\mu}_{[\,]} \qquad \mu_{ij} = \mu_{(ij)} + \mu_{[ij]} \tag{100}$$

The resultant force and moment about the origin on an arbitrary surface $S$ bounded by a circuit $C$ is

$$\int_S \mathbf{t}^{(n)} dS = \mathbf{F} \qquad \int_S t_i^{(n)} dS = F_i \tag{101}$$

$$\int_S \left[ \mathbf{r} \times \mathbf{t}^{(n)} + \mathbf{m}^{(n)} \right] dS = \mathbf{M}_O \qquad \int_S \left[ \varepsilon_{ijk} x_j t_k^{(n)} + m_i^{(n)} \right] dS = M_{Oi} \tag{102}$$

By using the relations (97) and (98) for tractions in terms of stresses, these also can be written as

$$\int_S d\mathbf{S} \bullet \boldsymbol{\sigma} = \mathbf{F} \qquad \int_S \sigma_{ji} n_j dS = F_i \tag{103}$$

$$\int_S d\mathbf{S} \bullet \left[ -\boldsymbol{\sigma} \times \mathbf{r} + \boldsymbol{\mu} \right] = \mathbf{M}_O \qquad \int_S \left[ \varepsilon_{ijk} x_j \sigma_{lk} + \mu_{li} \right] n_l dS = M_{Oi} \tag{104}$$

When the circuit $C$ shrinks to a point, the boundary surface $S$ becomes closed, representing the boundary surface of a part of the material continuum with volume $V$. Therefore, the resultant force and moment on this closed surface vanish, that is $\mathbf{F} = 0$ and $\mathbf{M}_O = 0$. By neglecting body forces, the force and moment balance governing equations for this part of the body enclosed by the surface $S$ are written, respectively, as

$$\int_S \mathbf{t}^{(n)} dS = 0 \qquad \int_S t_i^{(n)} dS = 0 \tag{105}$$

$$\int_S \left[ \mathbf{r} \times \mathbf{t}^{(n)} + \mathbf{m}^{(n)} \right] dS = 0 \qquad \int_S \left[ \varepsilon_{ijk} x_j t_k^{(n)} + m_i^{(n)} \right] dS = 0 \tag{106}$$

In terms of stresses, these become

$$\int_S d\mathbf{S} \bullet \boldsymbol{\sigma} = 0 \qquad \int_S \sigma_{ji} n_j dS = 0 \tag{107}$$

$$\int_S d\mathbf{S} \bullet [-\boldsymbol{\sigma} \times \mathbf{r} + \boldsymbol{\mu}] = 0 \qquad \int_S \left[ \sigma_{lk} \varepsilon_{kij} x_j + \mu_{li} \right] n_l dS = 0 \qquad (108)$$

By using the divergence theorem, the surface integrals are transformed to volume integral as

$$\int_V \nabla \bullet \boldsymbol{\sigma} \, dV = 0 \qquad \int_V \sigma_{ji,j} dV = 0 \qquad (109)$$

$$\int_V [\boldsymbol{\varepsilon} : \boldsymbol{\sigma} + \mathbf{r} \times \nabla \bullet \boldsymbol{\sigma} + \nabla \bullet \boldsymbol{\mu}] \, dV = 0 \qquad \int_V \left[ \varepsilon_{ijk} \sigma_{jk} + \varepsilon_{ijk} x_j \sigma_{lk,l} + \mu_{li,l} \right] dV = 0 \qquad (110)$$

By using the first equation (109), the second equation (110) reduces to

$$\int_V [\nabla \bullet \boldsymbol{\mu} + \boldsymbol{\varepsilon} : \boldsymbol{\sigma}] \, dV = 0 \qquad \int_V \left[ \mu_{ji,j} + \varepsilon_{ijk} \sigma_{jk} \right] dV = 0 \qquad (111)$$

By noticing the arbitrariness of $V$, we obtain the differential form of the equilibrium equations in the absence of body forces

$$\nabla \bullet \boldsymbol{\sigma} = 0 \qquad \sigma_{ji,j} = 0 \qquad (112)$$

$$\nabla \bullet \boldsymbol{\mu} + \boldsymbol{\varepsilon} : \boldsymbol{\sigma} = 0 \qquad \mu_{ji,j} + \varepsilon_{ijk} \sigma_{jk} = 0 \qquad (113)$$

Interestingly, we notice that the moment equation gives the skew-symmetric part of the force-stress tensor as

$$\boldsymbol{\sigma}_{[\,]} = -\frac{1}{2} \boldsymbol{\varepsilon} : (\nabla \bullet \boldsymbol{\mu}) \qquad \sigma_{[ij]} = -\frac{1}{2} \varepsilon_{ijk} \mu_{lk,l} \qquad (114)$$

### 5.2. Mindlin-Tiersten-Koiter couple stress theory

Mindlin and Tiersten (1962) and Koiter (1964) have shown that the displacement field $\mathbf{u}$ prescribed on a smooth part of the boundary $S$, specifies the normal component of the rotation $\omega^{(nn)} = \vec{\omega} \bullet \mathbf{n}$. Accordingly, they have demonstrated that material in a consistent couple stress theory does not support independent distributions of normal surface couple-traction $m^{(nn)} = \mathbf{m}^{(n)} \bullet \mathbf{n}$. This means

$$m^{(nn)} = \mathbf{m}^{(n)} \bullet \mathbf{n} = \mathbf{n} \bullet \boldsymbol{\mu} \bullet \mathbf{n} = 0 \qquad m^{(nn)} = m_i^{(n)} n_i = \mu_{ji} n_i n_j = 0 \tag{115}$$

Consequently, Mindlin and Tiersten (1962) and Koiter (1964) correctly established that five geometrical and five mechanical boundary conditions could be specified on a smooth surface. However, they did not realize the fundamental implication of (115) on the character of the couple-stress tensor.

The principle of virtual work can be developed by first multiplying the force and moment equilibrium equations (112) and (113) by energy conjugate virtual displacement $\delta \mathbf{u}$ and virtual rotation $\delta \vec{\boldsymbol{\omega}}$ and then integrating over some volume $V$. By noticing that $\delta \vec{\boldsymbol{\omega}} = \frac{1}{2} \nabla \times \delta \mathbf{u}$, after some manipulation, the principle of virtual work for this formulation (Hadjesfandiari and Dargush, 2011, 2015b) is written:

$$\int_S \left[ \mathbf{t}^{(n)} \bullet \delta \mathbf{u} + \mathbf{m}^{(n)} \bullet \delta \vec{\boldsymbol{\omega}} \right] dS = \int_V \left[ \boldsymbol{\sigma} : \delta \mathbf{e} + \boldsymbol{\mu} : \delta \mathbf{k} \right] dV \tag{116}$$

$$\int_S \left[ t_i^{(n)} \delta u_i + m_i^{(n)} \delta \omega_i \right] dS = \int_V \left[ \sigma_{ij} \delta e_{ij} + \mu_{ij} \delta k_{ij} \right] dV$$

Since there is no restriction on the form of the couple-stress tensor $\boldsymbol{\mu}$ in the original Mindlin-Tiersten-Koiter couple stress theory (MTK-CST), the bend-twist tensor $\mathbf{k} = \nabla \vec{\boldsymbol{\omega}}$ is considered as its energetically conjugate curvature tensor measure of deformation. This has some dramatic consequences. In particular, the original MTK-CST theory suffers from serious inconsistencies and difficulties with the underlying formulation (Eringen, 1968; Hadjesfandiari and Dargush, 2011, 2015a,b, 2016), which may be summarized as follows:

1. The inconsistency in boundary conditions, since the normal component of the couple-traction vector $m^{(nn)} = \mathbf{m}^{(n)} \bullet \mathbf{n}$ appears in the formulation violating the condition (115);

2. The appearance of an indeterminate spherical part in the couple-stress tensor, and thus, in the skew-symmetric part of the force-stress tensor;

3. The skew-symmetric tensor $\boldsymbol{\sigma}_{[\,]}$ should be derived from curl of a true vector or a second order skew-symmetric pseudo tensor, not from curl of a general couple-stress tensor $\boldsymbol{\mu}$;

4. The disturbing appearance of the body couple in the constitutive relation for the force-stress tensor;

5. The appearance of too many constitutive coefficients in the general anisotropic case, which makes MTK-CST less attractive from a practical perspective.

Unfortunately, Mindlin and Tiersten (1962) and Koiter (1964) confused the matter further by introducing the approximate reduced boundary condition method for $m^{(nn)} = \mathbf{m}^{(n)} \bullet \mathbf{n}$ from structural mechanics. However, we notice that there is a fundamental difference between couple stress theory and structural mechanics (Hadjesfandiari and Dargush, 2011, 2015b). The boundary condition (115) must be satisfied systematically in the formulation, because the couple stress theory is a continuum mechanics theory not an approximated structural mechanics. Nevertheless, this fundamental difficulty with boundary condition (115) and its impact on the formulation was not appreciated for a long time.

Eringen (1968) realized the indeterminacy problem as a major mathematical inconsistency in the original Mindlin-Tiersten-Koiter couple stress theory (MTK-CST), which he afterwards called indeterminate couple stress theory.

### 5.3. Consistent couple stress theory

After almost half a century of confusion created by the indeterminacy of MTK-CST, Hadjesfandiari and Dargush (2011) and Hadjesfandiari et al. (2015) discovered the consistent couple stress theory (C-CST). This theory not only answers the criticism of Eringen about the indeterminacy, but also resolves other inconsistencies in the original MTK-CST. The main achievement of this development is discovering the subtle skew-symmetric character of the couple-stress tensor

$$\boldsymbol{\mu}^t = -\boldsymbol{\mu} \qquad \mu_{ji} = -\mu_{ij} \qquad (117)$$

The fundamental step in this development is satisfying the requirement (115) that the normal component of the couple-traction vector must vanish on any arbitrary boundary surface in a systematic way, i.e. $m^{(nn)} = \mathbf{n} \bullet \boldsymbol{\mu} \bullet \mathbf{n} = 0$. This is what Mindlin, Tiersten and Koiter missed in their important developments of MTK-CST, although they correctly established the consistent set of boundary conditions. They did not realize that their consistent boundary conditions simply show the existence of the normal twisting couple-traction $m^{(nn)} = \mu_{ji} n_i n_j$ is physically impossible.

Interestingly, the principle of virtual work (116) for C-CST becomes

$$\int_S \left[ \mathbf{t}^{(n)} \bullet \delta \mathbf{u} + \mathbf{m}^{(n)} \bullet \delta \vec{\boldsymbol{\omega}} \right] dS = \int_V \left[ \boldsymbol{\sigma} : \delta \mathbf{e} + \boldsymbol{\mu} : \boldsymbol{\kappa} \right] dV \tag{118}$$

$$\int_S \left[ t_i^{(n)} \delta u_i + m_i^{(n)} \delta \omega_i \right] dS = \int_V \left[ \sigma_{ij} \delta e_{ij} + \mu_{ij} \delta \kappa_{ij} \right] dV$$

This shows that in C-CST the mean curvature tensor $\boldsymbol{\kappa}$ defined in (52), which is the skew-symmetric part of the bend-twist tensor $\mathbf{k}$, is the consistent curvature tensor measure of deformation. Therefore, the mean curvature tensor is energetically conjugate to the skew-symmetric couple-stress tensor $\boldsymbol{\mu}$. The skew-symmetric tensor $\boldsymbol{\mu}$ can also be represented by its dual true (polar) couple-stress vector $\vec{\boldsymbol{\mu}}$ (Hadjesfandiari and Dargush, 2011; Hadjesfandiari et al., 2015), where

$$\vec{\boldsymbol{\mu}} = \frac{1}{2} \boldsymbol{\varepsilon} : \boldsymbol{\mu}, \qquad \mu_i = \frac{1}{2} \varepsilon_{ijk} \mu_{jk} \tag{119}$$

$$\boldsymbol{\mu} = \boldsymbol{\varepsilon} \bullet \vec{\boldsymbol{\mu}}, \qquad \mu_{ij} = \varepsilon_{ijk} \mu_k \tag{120}$$

Consequently, the surface couple-traction vector $\mathbf{m}^{(n)}$ can be written as

$$\mathbf{m}^{(n)} = \mathbf{n} \bullet \boldsymbol{\mu} = \vec{\boldsymbol{\mu}} \times \mathbf{n} \qquad m_i^{(n)} = \mu_{ji} n_j = \varepsilon_{jik} \mu_k n_j \tag{121}$$

Since this traction is tangent to the surface, it obviously creates bending deformation.

Interestingly, we notice that the true mean curvature vector $\vec{\kappa}$ is energetically conjugate to the true couple-stress tensor $\vec{\mu}$. Therefore, the consistent skew-symmetric couple stress theory (C-CST) may be called the *vector couple stress theory*.

We also notice the remarkable relation

$$\boldsymbol{\mu} : \delta\boldsymbol{\kappa} = 2\vec{\boldsymbol{\mu}} \bullet \delta\vec{\boldsymbol{\kappa}} \qquad \mu_{ij}\delta\kappa_{ij} = 2\mu_i \delta\kappa_i \tag{122}$$

Therefore, the principle of virtual work for C-CST also can be written as

$$\int_S \left[ \mathbf{t}^{(n)} \bullet \delta\mathbf{u} + \mathbf{m}^{(n)} \bullet \delta\vec{\boldsymbol{\omega}} \right] dS = \int_V \left[ \boldsymbol{\sigma} : \delta\mathbf{e} + 2\vec{\boldsymbol{\mu}} \bullet \delta\vec{\boldsymbol{\kappa}} \right] dV \tag{123}$$

$$\int_S \left[ t_i^{(n)} \delta u_i + m_i^{(n)} \delta\omega_i \right] dS = \int_V \left[ \sigma_{ji} \delta e_{ij} + 2\mu_i \delta\kappa_i \right] dV$$

It is astounding to note that the skew-symmetric character of the couple-stress tensor immediately resolves the indeterminacy problem by establishing that there is no spherical component. As a result, the couple-stress tensor is determinate in the skew-symmetric C-CST. Interestingly, this shows that there is an interrelationship between the consistent mechanical boundary condition (115), $m^{(nn)} = 0$, and the determinacy of the couple-stress tensor; resolving one, resolves the other.

As we notice, in C-CST, the moment equations (113) give the skew-symmetric part of the force-stress tensor as

$$\boldsymbol{\sigma}_{[\,]} = \frac{1}{2}(\nabla\vec{\boldsymbol{\mu}} - \vec{\boldsymbol{\mu}}\nabla) \qquad \sigma_{[ij]} = \mu_{[j,i]} = \frac{1}{2}\left(\mu_{j,i} - \mu_{i,j}\right) \tag{124}$$

This relation can be elaborated if we introduce the pseudo (axial) vector $\mathbf{s}$ dual to the skew-symmetric part of the force-stress tensor $\boldsymbol{\sigma}_{[\,]}$, where

$$\mathbf{s} = \frac{1}{2}\boldsymbol{\varepsilon} : \boldsymbol{\sigma} \qquad s_i = \frac{1}{2}\varepsilon_{ijk}\sigma_{jk} \tag{125}$$

One may also write the dual relation

$$\boldsymbol{\sigma}_{[\,]} = \boldsymbol{\varepsilon} \bullet \mathbf{s} \qquad \sigma_{[ij]} = \varepsilon_{ijk} s_k \tag{126}$$

Thus, for the total force-stress tensor, we have

$$\boldsymbol{\sigma} = \boldsymbol{\sigma}_{(\ )} + \frac{1}{2}\left(\nabla\vec{\boldsymbol{\mu}} - \vec{\boldsymbol{\mu}}\nabla\right) \qquad \sigma_{ij} = \sigma_{(ij)} + \frac{1}{2}\left(\mu_{j,i} - \mu_{i,j}\right) \qquad (127)$$

As a result, the force governing equation (112) reduces to

$$\nabla \bullet \left\{\boldsymbol{\sigma}_{(\ )} + \frac{1}{2}\left(\nabla\vec{\boldsymbol{\mu}} - \vec{\boldsymbol{\mu}}\nabla\right)\right\} = 0 \qquad \left[\sigma_{(ji)} + \frac{1}{2}\left(\mu_{j,i} - \mu_{i,j}\right)\right]_{,j} = 0 \qquad (128)$$

It is important to note that C-CST answers the criticism of Eringen about the indeterminacy of the couple-stress tensor in MTK-CST without adding any new artificial law. It is remarkable that C-CST systematically links efforts of the Cosserats, Mindlin, Tiersten and Koiter and others in a span of a century. In Section 8, we will show that C-CST also enables us to resolve all the ambiguities and inconsistencies in the existing continuous defect theory to permit development of C-CDT.

In this paper, we have tried to be consistent with definitions and notations used by Mindlin and Tiersten (1962) and deWit (1970, 1973a-c), as much as possible. As a result, there are some discrepancies with our original definitions and notations in Hadjesfandiari and Dargush (2011, 2015a,b). The following highlight the main differences and similarities:

- The tensors $\boldsymbol{\omega}$ and $\boldsymbol{\kappa}$ defined here are negative of their corresponding definitions in Hadjesfandiari and Dargush (2011, 2015a,b);
- The vectors $\vec{\boldsymbol{\omega}}$ and $\vec{\boldsymbol{\kappa}}$ defined here are the same as their corresponding definitions in Hadjesfandiari and Dargush (2011, 2015a,b);
- The tensor $\boldsymbol{\mu}$ defined here is the same as its corresponding definition in Hadjesfandiari and Dargush (2011, 2015a,b);
- The vector $\vec{\boldsymbol{\mu}}$ defined here is negative of its corresponding definition in Hadjesfandiari and Dargush (2011, 2015);

As a result, the virtual work associated with couple-stress in (122) and (123) is $2\vec{\boldsymbol{\mu}} \bullet \delta\vec{\boldsymbol{\kappa}}$ in contrast to $-2\vec{\boldsymbol{\mu}} \bullet \delta\vec{\boldsymbol{\kappa}}$, as presented in Hadjesfandiari and Dargush (2011, 2015).

## 6. Original continuous defect theory

Here we present the original continuous defect theory (CDT) based on the work of Anthony (1970) and deWit (1970, 1973a-c). This theory plays a fundamental role in developing consistent continuous defect theory (C-CDT) in Section 8.

### 6.1. Elastic-plastic deformation, disclination and dislocation density tensors

When a body undergoes elastic-plastic deformation, the plastic part of the deformation specifies the defect content of the body. We should emphasize that the total deformation is still compatible, which means it satisfies the compatibility conditions (65) and (66) as long as the body deforms without breaking.

Here we assume the body is simply connected. Therefore, the rotations $\vec{\omega}$ and displacements $\mathbf{u}$ are single-valued functions. We notice that for this simply connected body undergoing arbitrary plastic deformation, some kinematical quantities can be decomposed into elastic and plastic parts. For example, the total strain $\mathbf{e}$ field can be decomposed into elastic $\mathbf{e}^E$ and plastic $\mathbf{e}^P$ parts, where

$$\mathbf{e} = \mathbf{e}^E + \mathbf{e}^P \qquad\qquad e_{ij} = e_{ij}^E + e_{ij}^P \tag{129}$$

On the other hand, the displacement vector field $\mathbf{u}$ represents the translational degrees of freedom at each point. Therefore, decomposing it into elastic $\mathbf{u}^E$ and plastic $\mathbf{u}^P$ parts is meaningless, that is, we should not write

$$\cancel{\mathbf{u} = \mathbf{u}^E + \mathbf{u}^P} \qquad\qquad \cancel{u_i = u_i^E + u_i^P} \tag{130}$$

Degrees of freedom cannot be decomposed into elastic and plastic parts.

The total strain $\mathbf{e}$ satisfies the compatibility condition (31), which now can be written as

$$\nabla \times \left(\mathbf{e}^E + \mathbf{e}^P\right) \times \nabla = 0 \qquad\qquad -\varepsilon_{ikl}\varepsilon_{jmn}\left(e_{ln,km}^E + e_{ln,km}^P\right) = 0 \tag{131}$$

Although the total strain tensor $\mathbf{e}$ is compatible, its elastic $\mathbf{e}^E$ and plastic $\mathbf{e}^P$ parts are not usually separately compatible. This means $\mathbf{e}^E$ and plastic $\mathbf{e}^P$ do not generally each satisfy the compatibility equation (31). This allows us to define the Saint-Venant's or strain incompatibility tensor $\mathbf{\eta}$ as

$$\mathbf{\eta} = \nabla \times \mathbf{e}^E \times \nabla = -\nabla \times \mathbf{e}^P \times \nabla \qquad \eta_{ij} = -\varepsilon_{ikl}\varepsilon_{jmn}e^E_{ln,km} = \varepsilon_{ikl}\varepsilon_{jmn}e^P_{ln,km} \qquad (132)$$

This symmetric true tensor measures the deviation from compatibility for incompatible elastic $\mathbf{e}^E$ and plastic $\mathbf{e}^P$ strains. This relation is obviously consistent with the fact that there is no general decomposition for the displacement as $\mathbf{u} = \mathbf{u}^E + \mathbf{u}^P$. The continuity equation for incompatibility tensor $\mathbf{\eta}$ follows from its definition (132) as

$$\nabla \bullet \mathbf{\eta} = 0 \qquad \eta_{ji,j} = 0 \qquad (133)$$

By some manipulation, the incompatibility tensor can also be expressed as (deWit, 1973a,b)

$$\mathbf{\eta} = \nabla^2 \mathbf{e}^E + \nabla\left[tr\left(\mathbf{e}^E\right)\right]\nabla - \nabla\left(\mathbf{e}^E \bullet \nabla\right) - \left(\nabla \bullet \mathbf{e}^E\right)\nabla - \left\{\nabla^2\left[tr\left(\mathbf{e}^E\right)\right] - \nabla \bullet \left(\nabla \bullet \mathbf{e}^E\right)\right\}\mathbf{I} \qquad (134)$$

$$\eta_{ij} = e^E_{ij,kk} + e^E_{kk,ij} - e^E_{jk,ik} - e^E_{ik,jk} - \delta_{ij}\left(e^E_{kk,ll} - e^E_{kl,kl}\right)$$

$$\mathbf{\eta} = -\nabla^2 \mathbf{e}^P - \nabla\left[tr\left(\mathbf{e}^P\right)\right]\nabla + \nabla\left(\mathbf{e}^P \bullet \nabla\right) + \left(\nabla \bullet \mathbf{e}^P\right)\nabla + \left\{\nabla^2\left[tr\left(\mathbf{e}^P\right)\right] - \nabla \bullet \left(\nabla \bullet \mathbf{e}^P\right)\right\}\mathbf{I} \qquad (135)$$

$$\eta_{ij} = -e^P_{ij,kk} - e^P_{kk,ij} + e^P_{jk,ik} + e^P_{ik,jk} + \delta_{ij}\left(e^P_{kk,ll} - e^P_{kl,kl}\right)$$

From this form of the incompatibility, we can derive the interesting relations

$$tr(\mathbf{\eta}) = \nabla^2\left[tr\left(\mathbf{e}^P\right)\right] - \nabla \bullet \left(\nabla \bullet \mathbf{e}^P\right) \qquad \eta_{kk} = e^P_{kk,ll} - e^P_{kl,kl} \qquad (136)$$

$$\mathbf{\eta} - \left[tr(\mathbf{\eta})\right]\mathbf{I} = -\nabla^2 \mathbf{e}^P - \nabla\left[tr\left(\mathbf{e}^P\right)\right]\nabla + \nabla\left(\mathbf{e}^P \bullet \nabla\right) + \left(\nabla \bullet \mathbf{e}^P\right)\nabla \qquad (137)$$

$$\eta_{ij} - \eta_{kk}\delta_{ij} = -e^P_{ij,kk} - e^P_{kk,ij} + e^P_{jk,ik} + e^P_{ik,jk}$$

Although the strain tensor $\mathbf{e}$ can be decomposed into elastic $\mathbf{e}^E$ and plastic $\mathbf{e}^P$ parts, the distortion tensor $\boldsymbol{\beta} = \nabla \mathbf{u}$ cannot. This is because the skew symmetric part of the distortion tensor $\boldsymbol{\beta} = \nabla \mathbf{u} = \mathbf{e} + \boldsymbol{\omega}$ represents the rotational degrees of freedom $\boldsymbol{\omega}$ of infinitesimal elements of matter at each point. This means we cannot decompose the rotation vector $\vec{\omega}$ into elastic $\vec{\omega}^E$ and plastic $\vec{\omega}^P$ parts, in a meaningful manner. Thus, we should not write

$$\vec{\omega} = \vec{\omega}^E + \vec{\omega}^P \qquad \omega_i = \omega_i^E + \omega_i^P \qquad (138)$$

This clearly shows that the distortion tensor $\boldsymbol{\beta} = \mathbf{e} + \boldsymbol{\omega}$ cannot be decomposed directly into elastic $\boldsymbol{\beta}^E$ and plastic $\boldsymbol{\beta}^P$ parts, where

$$\boldsymbol{\beta} = \boldsymbol{\beta}^E + \boldsymbol{\beta}^P \qquad \beta_i = \beta_i^E + \beta_i^P \qquad (139)$$

However, we notice that there is no restriction on the total bend-twist tensor $\mathbf{k} = \nabla \vec{\omega}$, which can be decomposed into elastic $\mathbf{k}^E$ and plastic $\mathbf{k}^P$ parts as

$$\mathbf{k} = \mathbf{k}^P + \mathbf{k}^E \qquad k_{ij} = k_{ij}^P + k_{ij}^E \qquad (140)$$

This is because the bend-twist tensor $\mathbf{k} = \nabla \vec{\omega} = \boldsymbol{\chi} + \boldsymbol{\kappa}$, its symmetric torsion tensor part $\boldsymbol{\chi}$ and the skew-symmetric mean curvature tensor part $\boldsymbol{\kappa}$, do not define any degrees of freedom.

The total bend-twist tensor $\mathbf{k}$ satisfies the first compatibility condition (65), which now can be written as

$$\nabla \times \left( \mathbf{k}^P + \mathbf{k}^E \right) = 0 \qquad \varepsilon_{ikl} \left( k_{lj,k}^E + k_{lj,k}^P \right) = 0 \qquad (141)$$

It is obvious that the elastic $\mathbf{k}^E$ and plastic $\mathbf{k}^P$ parts are usually incompatible, which means the individual parts do not generally satisfy the compatibility condition (65). In other words, $\mathbf{k}^E$ and $\mathbf{k}^P$ are not integrable functions to obtain $\mathbf{e}^E$ and $\mathbf{e}^P$. Interestingly, this allows us to define the bend-twist incompatibility or disclination density tensor $\boldsymbol{\theta}$, as

$$\boldsymbol{\theta} = \nabla \times \mathbf{k}^E = -\nabla \times \mathbf{k}^P \qquad \theta_{ij} = \varepsilon_{ikl} k_{lj,k}^E = -\varepsilon_{ikl} k_{lj,k}^P \qquad (142)$$

This true (polar) tensor measures the deviation from compatibility for the incompatible elastic $\mathbf{k}^E$ and plastic $\mathbf{k}^P$ bend-twist tensors.

The continuity equation for the generally non-symmetric true (polar) disclination density tensor $\boldsymbol{\theta}$ is obtained from its definition (142) as

$$\nabla \bullet \boldsymbol{\theta} = 0 \qquad \theta_{ji,j} = 0 \qquad (143)$$

As demonstrated previously in (44), the compatibility condition for the total distortion tensor $\boldsymbol{\beta} = \mathbf{e} + \boldsymbol{\omega}$ reduces to

$$\nabla \times \mathbf{e} + \boldsymbol{\varepsilon} : (\mathbf{k} \bullet \boldsymbol{\varepsilon}) = 0 \qquad \varepsilon_{ikl} e_{lj,k} + \varepsilon_{ikl} \varepsilon_{ljm} k_{km} = 0 \qquad (144)$$

which also can be written as

$$\boldsymbol{\varepsilon} : (\nabla \mathbf{e} + \mathbf{k} \bullet \boldsymbol{\varepsilon}) = 0 \qquad \varepsilon_{ikl} \left( e_{lj,k} + \varepsilon_{ljm} k_{km} \right) = 0 \qquad (145)$$

By decomposing the total strain and bend-twist tensors into their corresponding elastic and plastic parts, this relation can be written as

$$\boldsymbol{\varepsilon} : \left( \nabla \mathbf{e}^E + \mathbf{k}^E \bullet \boldsymbol{\varepsilon} \right) + \boldsymbol{\varepsilon} : \left( \nabla \mathbf{e}^P + \mathbf{k}^P \bullet \boldsymbol{\varepsilon} \right) = 0 \qquad (146)$$

$$\varepsilon_{ikl} \left( e^E_{lj,k} + \varepsilon_{ljm} k^E_{km} \right) + \varepsilon_{ikl} \left( e^P_{lj,k} + \varepsilon_{ljm} k^P_{km} \right) = 0$$

However, the elastic tensors $\mathbf{e}^E$ and $\mathbf{k}^E$, and plastic tensors $\mathbf{e}^P$ and $\mathbf{k}^P$ do not generally satisfy the compatibility condition (145) individually. Interestingly, this form allows us to define a distortion incompatibility or dislocation density pseudo (axial) tensor $\boldsymbol{\alpha}$, as

$$\boldsymbol{\alpha} = \boldsymbol{\varepsilon} : \left( \nabla \mathbf{e}^E + \mathbf{k}^E \bullet \boldsymbol{\varepsilon} \right) = -\boldsymbol{\varepsilon} : \left( \nabla \mathbf{e}^P + \mathbf{k}^P \bullet \boldsymbol{\varepsilon} \right) \qquad (147)$$

$$\alpha_{ij} = \varepsilon_{ikl} \left( e^E_{lj,k} + \varepsilon_{ljm} k^E_{km} \right) = -\varepsilon_{ikl} \left( e^P_{lj,k} + \varepsilon_{ljm} k^P_{km} \right)$$

By writing the compatibility relation for $\boldsymbol{\beta}$, repeated from (66), in the form

$$\nabla \times \mathbf{e} + tr(\mathbf{k}) \mathbf{I} - \mathbf{k}^t = 0 \qquad \varepsilon_{ikl} e_{lj,k} + k_{ll} \delta_{ij} - k_{ji} = 0 \qquad (148)$$

with

$$tr(\mathbf{k}) = 0 \qquad k_{ii} = 0 \qquad (149)$$

we notice

$$\nabla \times \left(\mathbf{e}^E + \mathbf{e}^P\right) + tr\left(\mathbf{k}^E + \mathbf{k}^P\right)\mathbf{I} - \mathbf{k}^{Et} - \mathbf{k}^{Pt} = 0 \qquad (150)$$

$$\varepsilon_{ikl}\left(e^E_{lj,k} + e^P_{lj,k}\right) + \left(k^E_{ll} + k^P_{ll}\right)\delta_{ij} - \left(k^E_{ji} + k^P_{ji}\right) = 0$$

$$tr\left(\mathbf{k}^E + \mathbf{k}^P\right) = 0 \qquad k^E_{ii} + k^P_{ii} = 0 \qquad (151)$$

As a result, we obtain the other form of the dislocation density $\boldsymbol{\alpha}$ as

$$\boldsymbol{\alpha} = \nabla \times \mathbf{e}^E + tr\left(\mathbf{k}^E\right)\mathbf{I} - \mathbf{k}^{Et} = -\left[\nabla \times \mathbf{e}^P + tr\left(\mathbf{k}^P\right)\mathbf{I} - \mathbf{k}^{Pt}\right] \qquad (152)$$

$$\alpha_{ij} = \varepsilon_{ikl}e^E_{lj,k} + k^E_{ll}\delta_{ij} - k^E_{ji} = -\varepsilon_{ikl}e^P_{lj,k} - k^P_{ll}\delta_{ij} + k^P_{ji}$$

where

$$tr\left(\mathbf{k}^E\right) = -tr\left(\mathbf{k}^P\right) \qquad k^E_{ii} = -k^P_{ii} \qquad (153)$$

In this paper, we use both equivalent forms of the dislocation tensor $\boldsymbol{\alpha}$. Thus,

$$\boldsymbol{\alpha} = -\boldsymbol{\varepsilon} : \left(\nabla \mathbf{e}^P + \mathbf{k}^P \bullet \boldsymbol{\varepsilon}\right) \qquad \alpha_{ij} = -\varepsilon_{ikl}\left(e^P_{lj,k} + \varepsilon_{ljm}k^P_{km}\right) \qquad (154)$$

$$\boldsymbol{\alpha} = -\nabla \times \mathbf{e}^P - tr\left(\mathbf{k}^P\right)\mathbf{I} + \mathbf{k}^{Pt} \qquad \alpha_{ij} = -\varepsilon_{ikl}e^P_{lj,k} - k^P_{ll}\delta_{ij} + k^P_{ji} \qquad (155)$$

It should be emphasized that since there is no decomposition of distortion as $\boldsymbol{\beta} = \boldsymbol{\beta}^E + \boldsymbol{\beta}^P$, it is meaningless to define the dislocation tensor $\boldsymbol{\alpha}$ as

$$\boldsymbol{\alpha} = \nabla \times \boldsymbol{\beta}^E = -\nabla \times \boldsymbol{\beta}^P \qquad (156)$$

However, we notice that this expression is what Nabarro (1967) and deWit (1970) have defined as the dislocation tensor $\boldsymbol{\alpha}$.

The transpose of the dislocation pseudo tensor becomes

$$\boldsymbol{\alpha}^t = \mathbf{e}^P \times \nabla - tr\left(\mathbf{k}^P\right)\mathbf{I} + \mathbf{k}^P \qquad (157)$$

The curl of this relation is

$$\nabla \times \boldsymbol{\alpha}^t = \nabla \times \mathbf{e}^P \times \nabla - \nabla \times tr\left(\mathbf{k}^P\right)\mathbf{I} + \nabla \times \mathbf{k}^P \qquad (158)$$

Interestingly, by noticing that $\boldsymbol{\eta} = -\nabla \times \mathbf{e}^P \times \nabla$ and $\boldsymbol{\theta} = -\nabla \times \mathbf{k}^P$, we obtain the relation for the disclination tensor as

$$\boldsymbol{\theta} = -\nabla \times \boldsymbol{\alpha}^t - \boldsymbol{\eta} + \boldsymbol{\varepsilon} \bullet \nabla \left[ tr\left(\mathbf{k}^P\right) \right] \qquad \theta_{ij} = -\varepsilon_{ikl}\alpha_{jl,k} - \eta_{ij} + \varepsilon_{ijk}k^P_{ll,k} \qquad (159)$$

Now let us derive the continuity relation for the dislocation density tensor $\boldsymbol{\alpha}$ by obtaining its divergence $\nabla \bullet \boldsymbol{\alpha} = \alpha_{ji,j}\mathbf{i}_i$, where

$$\alpha_{ji} = -\varepsilon_{jkl}\left(e^P_{li,k} + \varepsilon_{lim}k^P_{km}\right) \qquad (160)$$

It follows that

$$\begin{aligned}
\alpha_{ji,j} &= -\varepsilon_{jkl}\left(e^P_{li,kj} + \varepsilon_{lim}k^P_{km,j}\right) \\
&= -\varepsilon_{jkl}\varepsilon_{lim}k^P_{km,j} \\
&= -\varepsilon_{lim}\varepsilon_{jkl}k^P_{km,j} \\
&= -\varepsilon_{lim}\varepsilon_{ljk}k^P_{km,j}
\end{aligned} \qquad (161)$$

By noticing $\theta_{lm} = -\varepsilon_{ljk}k^P_{km,j}$, this can be written as

$$\alpha_{ji,j} + \varepsilon_{ilm}\theta_{lm} = 0 \qquad (162)$$

Therefore, the continuity conditions for disclination and dislocation density tensors are

$$\nabla \bullet \boldsymbol{\theta} = 0 \qquad\qquad \theta_{ji,j} = 0 \qquad (163)$$

$$\nabla \bullet \boldsymbol{\alpha} + \boldsymbol{\varepsilon}:\boldsymbol{\theta} = 0 \qquad\qquad \alpha_{ji,j} + \varepsilon_{ijk}\theta_{jk} = 0 \qquad (164)$$

which have been presented by Anthony (1970) and deWit (1970, 1973a-c). Notice the similarity of (163) and (164) with the governing equilibrium equations from the statical couple stress theory (112) and (113).

The disclination and dislocation density tensors $\boldsymbol{\theta}$ and $\boldsymbol{\alpha}$ in these relations are generally non-symmetric. As a result, they can be decomposed into symmetric and skew-symmetric parts as

$$\boldsymbol{\theta} = \boldsymbol{\theta}_{(\ )} + \boldsymbol{\theta}_{[\ ]} \qquad\qquad \theta_{ij} = \theta_{(ij)} + \theta_{[ij]} \qquad (165)$$

$$\boldsymbol{\alpha} = \boldsymbol{\alpha}_{(\ )} + \boldsymbol{\alpha}_{[\ ]} \qquad\qquad \alpha_{ij} = \alpha_{(ij)} + \alpha_{[ij]} \qquad (166)$$

We notice that the second continuity equation (164) gives the skew-symmetric part of the disclination density tensor $\boldsymbol{\theta}_{[\ ]}$ as

$$\boldsymbol{\theta}_{[\ ]} = \frac{1}{2}\boldsymbol{\varepsilon}\bullet(\nabla\bullet\boldsymbol{\alpha}) \qquad \theta_{[ij]} = -\frac{1}{2}\varepsilon_{ijk}\alpha_{lk,l} \qquad (167)$$

Thus, for the total disclination tensor, we have

$$\boldsymbol{\theta} = \boldsymbol{\theta}_{(\ )} + \frac{1}{2}\boldsymbol{\varepsilon}\bullet(\nabla\bullet\boldsymbol{\alpha}) \qquad \theta_{ij} = \theta_{(ij)} - \frac{1}{2}\varepsilon_{ijk}\alpha_{lk,l} \qquad (168)$$

The skew-symmetric part $\boldsymbol{\theta}_{[\ ]}$ is called the twist disclination density tensor and denoted by $\boldsymbol{\Theta} = \boldsymbol{\theta}_{[\ ]}$. We notice that relation (167) shows that this tensor is completely defined by the dislocation density tensor $\boldsymbol{\alpha}$. This skew-symmetric tensor can also be represented by its dual pseudo (axial) twist disclination tensor vector $\vec{\boldsymbol{\Theta}} = \vec{\boldsymbol{\theta}}_{[\ ]}$, where

$$\vec{\boldsymbol{\Theta}} = \frac{1}{2}\boldsymbol{\varepsilon}:\boldsymbol{\theta} \qquad \Theta_i = \frac{1}{2}\varepsilon_{ijk}\theta_{jk} \qquad (169)$$

$$\boldsymbol{\Theta} = \boldsymbol{\varepsilon}\bullet\vec{\boldsymbol{\Theta}} \qquad \Theta_{ij} = \varepsilon_{ijk}\Theta_k \qquad (170)$$

### 6.2. Weingarten's theorem and continuous defect theory

The disclination and dislocation density tensors $\boldsymbol{\theta}$ and $\boldsymbol{\alpha}$ have been defined as the negative of the incompatibility of the plastic part of the bend-twist tensor $\mathbf{k}$ (142) and the reduced distortion tensor $\boldsymbol{\beta}$ in the forms (154) and (155). Here by using Weingarten's theorem for an arbitrary closed Burgers circuit $C$, we demonstrate that the disclination and dislocation density tensors $\boldsymbol{\theta}$ and $\boldsymbol{\alpha}$ are suitable measures of internal defects in an elastic-plastic body. For simplicity, the body is assumed to be simply connected.

For a simply connected body, the compatibility of deformation results in a single valued character of $\vec{\boldsymbol{\omega}}$ and $\mathbf{u}$. Hence, for any arbitrary closed Burgers circuit $C$, we have

$$\oint_C d\vec{\boldsymbol{\omega}} = 0 \qquad \oint_C d\omega_i = 0 \tag{171}$$

$$\oint_C d\mathbf{u} = 0 \qquad \oint_C du_i = 0 \tag{172}$$

Therefore, Weingarten's theorem for this simply connected body results in

$$\oint_C d\mathbf{r} \bullet \mathbf{k} = 0 \qquad \oint_C k_{ji} dx_j = 0 \tag{173}$$

$$\oint_C d\mathbf{r} \bullet [\mathbf{e} - \mathbf{k} \times \mathbf{r}] = 0 \qquad \oint_C \left(e_{ki} - \varepsilon_{ilj} k_{kl} x_j \right) dx_k = 0 \tag{174}$$

which means there is no discrete Frank and Burgers vectors in the simply connected body. By considering the elastic and plastic bend-twist tensors $\mathbf{k}^E$ and $\mathbf{k}^P$, along with the elastic and plastic strain tensors $\mathbf{e}^E$ and $\mathbf{e}^P$, these contour or circuit integrals can be expressed as

$$\oint_C d\mathbf{r} \bullet \mathbf{k}^E + \oint_C d\mathbf{r} \bullet \mathbf{k}^P = 0 \qquad \oint_C k_{ji}^E dx_j + \oint_C k_{ji}^P dx_j = 0 \tag{175}$$

$$\oint_C d\mathbf{r} \bullet \left[\mathbf{e}^E - \mathbf{k}^E \times \mathbf{r}\right] + \oint_C d\mathbf{r} \bullet \left[\mathbf{e}^P - \mathbf{k}^P \times \mathbf{r}\right] = 0 \tag{176}$$

$$\oint_C \left(e_{ki}^E - \varepsilon_{ilj} k_{kl}^E x_j \right) dx_k + \oint_C \left(e_{ki}^P - \varepsilon_{ilj} k_{kl}^P x_j \right) dx_k = 0$$

Now the individual elastic and plastic strain tensors $\mathbf{e}^E$ and $\mathbf{e}^P$, as well as elastic and plastic bend-twist tensors $\mathbf{k}^E$ and $\mathbf{k}^P$, are generally incompatible. Accordingly, this allows us to define the total internal Frank true vector $\vec{\Omega}$ and the total internal Burgers pseudo vector $\mathbf{B}_O$ corresponding to the plastic or elastic deformations for the circuit $C$ relative to the origin as

$$\vec{\Omega} = \oint_C d\mathbf{r} \bullet \mathbf{k}^E = -\oint_C d\mathbf{r} \bullet \mathbf{k}^P \qquad \Omega_i = \oint_C k_{ji}^E dx_j = -\oint_C k_{ji}^P dx_j \tag{177}$$

$$\mathbf{B}_O = \oint_C d\mathbf{r} \bullet \left[\mathbf{e}^E - \mathbf{k}^E \times \mathbf{r}\right] = -\oint_C d\mathbf{r} \bullet \left[\mathbf{e}^P - \mathbf{k}^P \times \mathbf{r}\right] \tag{178}$$

$$B_{Oi} = \oint_C \left(e_{ki}^E - \varepsilon_{ilj} k_{kl}^E x_j \right) dx_k = -\oint_C \left(e_{ki}^P - \varepsilon_{ilj} k_{kl}^P x_j \right) dx_k$$

However, these total internal Frank vector $\vec{\Omega}$ and Burgers vector $\mathbf{B}_O$ depend on the closed Burger circuit $C$. Since the total internal Frank vector is independent of the origin, it has been denoted by $\vec{\Omega}$ instead of $\vec{\Omega}_O$. We notice that the total internal Burgers vector $\mathbf{B}$ relative to an arbitrary point with position $\mathbf{r}$ is

$$\begin{aligned} \mathbf{B} &= \oint_C d\mathbf{r}' \bullet \left[ -\mathbf{e}'^P + \mathbf{k}'^P \times (\mathbf{r}' - \mathbf{r}) \right] \\ &= \oint_C d\mathbf{r}' \bullet \left[ -\mathbf{e}'^P + \mathbf{k}'^P \times \mathbf{r}' \right] - \left( \oint_C d\mathbf{r}' \bullet \mathbf{k}'^P \right) \times \mathbf{r} \end{aligned} \tag{179}$$

which can be written as

$$\begin{aligned} \mathbf{B} &= \mathbf{B}_O + \vec{\Omega} \times \mathbf{r} \\ &= \mathbf{B}_O + (-\mathbf{r}) \times \vec{\Omega} \end{aligned} \tag{180}$$

This confirms that for plastic induced defects, the internal Burgers vector $\mathbf{B}$ relative to the point at $\mathbf{r}$ is equal to the internal Burgers vector $\mathbf{B}_O$ at the origin, plus the moment of the internal Frank vector $\vec{\Omega}$ at the origin about this point.

Now we use Stokes' theorem to transform the circuit integrals (177) and (178) to surface integrals. By applying Stokes' theorem to any surface $S$ in the body bounded by the circuit $C$, the Frank vector $\vec{\Omega}$ in (177) becomes

$$\vec{\Omega} = -\int_S d\mathbf{S} \bullet \nabla \times \mathbf{k}^P \qquad \Omega_i = -\int_S \varepsilon_{jkl} k^P_{li,k} n_j dS \tag{181}$$

By noticing that the disclination density true tensor $\boldsymbol{\theta}$ is defined as

$$\boldsymbol{\theta} = -\nabla \times \mathbf{k}^P \qquad \theta_{ij} = -\varepsilon_{ikl} k^P_{lj,k} \tag{182}$$

the internal Frank vector can be written as

$$\vec{\Omega} = \int_S d\mathbf{S} \bullet \boldsymbol{\theta} \qquad \Omega_i = \int_S \theta_{ji} n_j dS \tag{183}$$

Similarly, by using Stokes' theorem for the internal Burgers vector $\mathbf{B}_O$ relative to the origin in (178), we obtain

$$\mathbf{B}_O = -\int_S d\mathbf{S} \bullet \nabla \times \left[ \mathbf{e}^P - \mathbf{k}^P \times \mathbf{r} \right] \qquad B_{Oi} = -\oint_C \varepsilon_{pqk} \left( e_{ki}^P - \varepsilon_{ilj} k_{kl}^P x_j \right)_{,q} n_p dS \qquad (184)$$

or

$$\mathbf{B}_O = -\int_S d\mathbf{S} \bullet \left[ \nabla \times \mathbf{e}^P - \nabla \times \left( \mathbf{k}^P \times \mathbf{r} \right) \right] \qquad B_{Oi} = -\int_S \left[ \varepsilon_{pqk} e_{ki,q}^P - \varepsilon_{pqk} \left( \varepsilon_{ilj} k_{kl}^P x_j \right)_{,q} \right] n_p dS \quad (185)$$

After some manipulation, this also can be written as

$$\mathbf{B}_O = \int_S d\mathbf{S} \bullet \left[ -\nabla \times \mathbf{e}^P - tr\left( \mathbf{k}^P \right) \mathbf{I} + \mathbf{k}^{Pt} + \left( \nabla \times \mathbf{k}^P \right) \times \mathbf{r} \right] \qquad (186)$$

$$B_{Oi} = \int_S \left[ -\varepsilon_{pqk} e_{ki,q}^P - k_{ll}^P \delta_{pi} + k_{ip}^P + \varepsilon_{ilj} \varepsilon_{pqk} k_{kl,q}^P x_j \right] n_p dS$$

Now by noticing the definition of the disclination density true tensor $\boldsymbol{\theta}$ in (182) and the dislocation tensor $\boldsymbol{\alpha}$ as

$$\boldsymbol{\alpha} = -\nabla \times \mathbf{e}^P - tr\left( \mathbf{k}^P \right) \mathbf{I} + \mathbf{k}^{Pt} \qquad \alpha_{ij} = -\varepsilon_{ikl} e_{lj,k}^P - k_{ll}^P \delta_{ij} + k_{ji}^P \qquad (187)$$

the total internal Burgers vector relative to the origin can be written as

$$\mathbf{B}_O = \int_S d\mathbf{S} \bullet \left[ \boldsymbol{\alpha} - \boldsymbol{\theta} \times \mathbf{r} \right] \qquad B_{Oi} = \int_S \left[ \alpha_{ki} + \varepsilon_{ijl} x_j \theta_{kl} \right] n_k dS \qquad (188)$$

Therefore, the total internal plastic induced Frank true vector $\vec{\Omega}$ and Burgers pseudo vector $\mathbf{B}_O$ relative to the origin for the surface $S$ bounded by circuit $C$ become

$$\vec{\Omega} = \int_S \mathbf{n} \bullet \boldsymbol{\theta} dS \qquad \Omega_i = \int_S \theta_{ji} n_j dS \qquad (189)$$

$$\mathbf{B}_O = \int_S \left[ \mathbf{n} \bullet \boldsymbol{\alpha} + \mathbf{r} \times \left( \mathbf{n} \bullet \boldsymbol{\theta} \right) \right] dS \qquad B_{Oi} = \int_S \left[ \alpha_{ji} n_j + \varepsilon_{ijl} x_j \theta_{kl} n_k \right] dS \qquad (190)$$

These relations show that $\vec{\Omega}$ is the Frank true vector of the plastic induced disclinations crossing $S$, and $\mathbf{B}_O$ is the general Burgers pseudo vector of the plastic induced dislocations and disclinations crossing that surface. Remarkably, Anthony (1970) and deWit (1970) noted that there is a dualism between this continuous defect theory (CDT) and the original Mindlin-Tiersten-Koiter couple stress theory (MTK-CST). Interestingly, Kondo (1968) has reduced the study of many physical phenomena, such as yielding, to a study of geometry in Schaefer space. This suggests that there is a duality between the geometry of defects and its statics in continuum mechanics (Anthony, 1970; deWit, 1970).

## 7. Original continuous defect theory with new perspective

Here we investigate the benefits and shortcomings of the original continuous defect theory (CDT). For this purpose, we introduce new additional concepts based on duality of its geometry and statics, which will be very helpful in developing consistent continuous defect theory (C-CDT) in Section 8. This also enables us to recognize the inconsistencies of the original CDT.

### 7.1. Original continuous defect theory as a dual couple stress theory

As mentioned, Anthony (1970) and deWit (1970) noted the fundamental dualism between the geometry of continuous defect theory (CDT) and the statics in original Mindlin-Tiersten-Koiter couple stress theory (MTK-CST). This enables us to introduce new concepts, which are helpful in studying defect theory analogous to couple stress theory.

In this analogous original continuous defect theory, it is postulated that the state of defect on a surface element $dS$ with unit normal vector $\mathbf{n}$ is specified by means of an infinitesimal Frank true vector $d\vec{\Omega}$ and an infinitesimal Burgers couple pseudo vector with Burgers vector $d\mathbf{b}$, where

$$d\vec{\Omega} = \mathbf{q}^{(n)} dS \qquad d\Omega_i = q_i^{(n)} dS \qquad (191)$$

$$d\mathbf{b} = \mathbf{a}^{(n)} dS \qquad db_i = a_i^{(n)} dS \qquad (192)$$

Here $\mathbf{q}^{(n)}$ denotes the Frank-traction true vector, while $\mathbf{a}^{(n)}$ stands for the Burgers-traction pseudo vector on the surface. More precisely, the latter could be defined as the Burgers couple-traction. However, for convenience, this will be called simply the Burgers-traction in the following.

The state of defect at each point is known, if the Frank-traction vector $\mathbf{q}^{(n)}$ and Burgers-traction vector $\mathbf{a}^{(n)}$ on arbitrary surfaces at that point are known. Interestingly, it is necessary to know only the Frank- and Burgers-tractions on three different planes passing the point. When these planes are taken parallel to the coordinate planes with unit normal $\mathbf{n}$ along the coordinate axes, the Frank-traction vectors are $\mathbf{q}^{(1)}$, $\mathbf{q}^{(2)}$ and $\mathbf{q}^{(3)}$, and the Burgers-traction vectors are $\mathbf{a}^{(1)}$, $\mathbf{a}^{(2)}$ and $\mathbf{a}^{(3)}$, where

$$\mathbf{q}^{(i)} = q_1^{(i)}\mathbf{i}_1 + q_2^{(i)}\mathbf{i}_2 + q_3^{(i)}\mathbf{i}_3 = q_j^{(i)}\mathbf{i}_j \tag{193}$$

and

$$\mathbf{a}^{(i)} = a_1^{(i)}\mathbf{i}_1 + a_2^{(i)}\mathbf{i}_2 + a_3^{(i)}\mathbf{i}_3 = a_j^{(i)}\mathbf{i}_j \tag{194}$$

We notice that the arrays of components of Frank-traction vectors

$$\begin{bmatrix} \mathbf{q}^{(1)} & \mathbf{q}^{(2)} & \mathbf{q}^{(3)} \end{bmatrix}^t = \begin{bmatrix} \mathbf{q}^{(1)t} \\ \mathbf{q}^{(2)t} \\ \mathbf{q}^{(3)t} \end{bmatrix} \tag{195}$$

and the arrays of components of Burgers-traction vectors

$$\begin{bmatrix} \mathbf{a}^{(1)} & \mathbf{a}^{(2)} & \mathbf{a}^{(3)} \end{bmatrix}^t = \begin{bmatrix} \mathbf{a}^{(1)t} \\ \mathbf{a}^{(2)t} \\ \mathbf{a}^{(3)t} \end{bmatrix} \tag{196}$$

represent the disclination and dislocation density tensors $\boldsymbol{\theta}$ and $\boldsymbol{\alpha}$, respectively, where

$$[\theta_{ij}] = \begin{bmatrix} \theta_{11} & \theta_{12} & \theta_{13} \\ \theta_{21} & \theta_{22} & \theta_{23} \\ \theta_{31} & \theta_{32} & \theta_{33} \end{bmatrix} = \begin{bmatrix} q_1^{(1)} & q_2^{(1)} & q_3^{(1)} \\ q_1^{(2)} & q_2^{(2)} & q_3^{(2)} \\ q_1^{(3)} & q_2^{(3)} & q_3^{(3)} \end{bmatrix} \tag{197}$$

$$[\alpha_{ij}] = \begin{bmatrix} \alpha_{11} & \alpha_{12} & \alpha_{13} \\ \alpha_{21} & \alpha_{22} & \alpha_{23} \\ \alpha_{31} & \alpha_{32} & \alpha_{33} \end{bmatrix} = \begin{bmatrix} a_1^{(1)} & a_2^{(1)} & a_3^{(1)} \\ a_1^{(2)} & a_2^{(2)} & a_3^{(2)} \\ a_1^{(3)} & a_2^{(3)} & a_3^{(3)} \end{bmatrix} \tag{198}$$

Therefore, the internal continuous defects are described by the generally non-symmetric true (polar) disclination density tensor $\boldsymbol{\theta}$ and non-symmetric pseudo (axial) dislocation density tensor $\boldsymbol{\alpha}$, where the Frank-traction true vector $\mathbf{q}^{(n)}$ and Burgers-traction pseudo vector $\mathbf{a}^{(n)}$ on any plane with unit normal vector $\mathbf{n}$ are represented by

$$\mathbf{q}^{(n)} = \mathbf{n} \bullet \boldsymbol{\theta} \qquad q_i^{(n)} = \theta_{ji} n_j \qquad (199)$$

$$\mathbf{a}^{(n)} = \mathbf{n} \bullet \boldsymbol{\alpha} \qquad a_i^{(n)} = \alpha_{ji} n_j \qquad (200)$$

The dualism between the geometry and statics of the original CDT in the framework of Mindlin-Tiersten-Koiter couple stress theory (MTK-CST) shows that the disclination and dislocation density tensors $\boldsymbol{\theta}$ and $\boldsymbol{\alpha}$ may be called the Frank-stress and Burgers couple-stress tensors, respectively. Again, for simplicity, we refer to the latter as the Burgers-stress tensor.

Interestingly, the six types of defect corresponding to the components of Frank-traction $\mathbf{q}^{(n)}$ and Burgers-traction $\mathbf{a}^{(n)}$ on the surface can be symbolically represented by the six discrete disclinations and dislocations shown in Figures 2 and 3, where the cut plane corresponds to the surface with unit normal vector $\mathbf{n}$. However, we notice that the total deformation in this continuous defect model is compatible and there is no jump or discontinuity.

For a surface $S$, which is bounded by the closed curve $C$, the total Frank true vector $\vec{\Omega}$ and the Burgers pseudo vector $\mathbf{B}_O$ relative to the origin are given as

$$\vec{\Omega} = \int_S \mathbf{q}^{(n)} dS \qquad \Omega_i = \int_S q_i^{(n)} dS \qquad (201)$$

$$\mathbf{B}_O = \int_S \left[ \mathbf{a}^{(n)} + \mathbf{r} \times \mathbf{q}^{(n)} \right] dS \qquad B_{Oi} = \int_S \left[ a_i^{(n)} + \varepsilon_{ijl} x_j q_l^{(n)} \right] dS \qquad (202)$$

By using the relations for Frank- and Burgers-tractions in terms of the disclination and dislocation densities $\boldsymbol{\theta}$ and $\boldsymbol{\alpha}$ (Frank- and Burgers-stresses), these relations become

$$\vec{\Omega} = \int_S d\mathbf{S} \bullet \boldsymbol{\theta} \qquad \Omega_i = \int_S \theta_{ji} n_j dS \qquad (203)$$

$$\mathbf{B}_O = \int_S d\mathbf{S} \bullet [\mathbf{\alpha} - \mathbf{\theta} \times \mathbf{r}] \qquad B_i = \int_S \left[\alpha_{ki} + \varepsilon_{ijl} x_j \theta_{kl}\right] n_k dS \qquad (204)$$

Let us investigate the character of these relations for the case when the surface $S$ is closed. We notice that when the circuit $C$ shrinks to a point, the boundary surface $S$ becomes closed specifying the boundary surface of an arbitrary volume $V$ of the body. We notice that for this closed surface the total Frank true vector and the Burgers pseudo vector relative to any point vanish, that is, $\vec{\Omega} = 0$ and $\mathbf{B}_O = 0$. Therefore, the Frank and Burgers governing continuity equations for this part of the body are written, respectively, as

$$\int_S \mathbf{q}^{(n)} dS = 0 \qquad \int_S q_i^{(n)} dS = 0 \qquad (205)$$

$$\int_S \left[\mathbf{a}^{(n)} + \mathbf{r} \times \mathbf{q}^{(n)}\right] dS = 0 \qquad \int_S \left[a_i^{(n)} + \varepsilon_{ijl} x_j q_l^{(n)}\right] dS = 0 \qquad (206)$$

In terms of the disclination and dislocation density (Frank- and Burgers-stress) tensors, these relations are

$$\int_S d\mathbf{S} \bullet \mathbf{\theta} = 0 \qquad \int_S \theta_{ji} n_j dS = 0 \qquad (207)$$

$$\int_S d\mathbf{S} \bullet [\mathbf{\alpha} - \mathbf{\theta} \times \mathbf{r}] = 0 \qquad \int_S \left[\alpha_{ki} + \varepsilon_{ijl} x_j \theta_{kl}\right] n_k dS = 0 \qquad (208)$$

By using the divergence theorem to transform the surface integrals (207) and (208) to volume integrals and noticing the arbitrariness of the volume $V$, we again can obtain the governing continuity equations for disclination and dislocation density tensors as

$$\nabla \bullet \mathbf{\theta} = 0 \qquad \theta_{ji,j} = 0 \qquad (209)$$

$$\nabla \bullet \mathbf{\alpha} + \mathbf{\varepsilon} : \mathbf{\theta} = 0 \qquad \alpha_{ji,j} + \varepsilon_{ijk} \theta_{jk} = 0 \qquad (210)$$

Notice that the steps are analogous to those in the Mindlin-Tiersten-Koiter couple stress theory (MTK-CST).

We have presented in this section the continuous defect theory (CDT) of Anthony (1970) and deWit (1970, 1973a-c) for a simply connected body with new perspectives. It turns out that this theory, although incomplete, will play a fundamental role in developing Consistent Continuous Defect Theory (C-CDT) in Section 8. It is very important to note that Weingarten's theorem has led to a continuous defect theory of disclinations and dislocations, which is dual to the statics of continuum mechanics, based on Mindlin-Tiersten-Koiter couple stress theory (MTK-CST). Table 1 summarizes the dualism between the geometry and statics of CDT in the original Mindlin-Tiersten-Koiter couple stress theory (deWit, 1970). Interestingly, deWit (1970) has also addressed the geometrical character of the force- and couple-stress tensors by using the Günther stress function tensors (Günther, 1958). We leave a discussion of this aspect to future work.

**Table 1. Dualism between geometry and statics in the original continuous defect theory (deWit, 1970)**

| Geometry<br>CDT | Statics<br>MTK-CST |
|---|---|
| Frank vector $\vec{\boldsymbol{\Omega}}$ | Force vector $\mathbf{F}$ |
| Burgers vector $\mathbf{B}$ | Moment vector $\mathbf{M}$ |
| Frank-traction vector $\mathbf{q}^{(n)}$ | Force-traction vector $\mathbf{t}^{(n)}$ |
| Burgers-traction vector $\mathbf{a}^{(n)}$<br>(Burgers couple-traction vector) | Couple-traction vector $\mathbf{m}^{(n)}$ |
| Disclination density tensor $\boldsymbol{\theta}$<br>(Frank-stress tensor) | Force-stress tensor $\boldsymbol{\sigma}$ |
| Dislocation density tensor $\boldsymbol{\alpha}$<br>(Burgers-stress tensor) | Couple-stress tensor $\boldsymbol{\mu}$ |
| Governing equations<br>(Continuity equations)<br>$\nabla \bullet \boldsymbol{\theta} = 0$<br>$\nabla \bullet \boldsymbol{\alpha} + \boldsymbol{\varepsilon} : \boldsymbol{\theta} = 0$ | Governing equations<br>(Equilibrium equations)<br>$\nabla \bullet \boldsymbol{\sigma} = 0$<br>$\nabla \bullet \boldsymbol{\mu} + \boldsymbol{\varepsilon} : \boldsymbol{\sigma} = 0$ |
| $\boldsymbol{\theta} = -\nabla \times \boldsymbol{\alpha}^t - \boldsymbol{\eta} + \boldsymbol{\varepsilon} \bullet \nabla \left[ tr\left(\mathbf{k}^P\right) \right]$<br>$\boldsymbol{\alpha} = -\nabla \times \mathbf{e}^P - tr\left(\mathbf{k}^P\right)\mathbf{I} + \mathbf{k}^{Pt}$<br>$\boldsymbol{\eta} = -\nabla \times \mathbf{e}^P \times \nabla$ | |

Now we investigate the character of external concentrated Frank and Burgers vectors in the body, which are analogous to concentrated force and couple force (moment) vectors in the body, respectively. Let us assume the body has a cavity, where the outer surface is $S_1$ and inner surface is $S_2$. Note that the body with cavities (e.g., spherical cavities) is still simply connected. This means the relations (205) and (206) are still valid, where $S = S_1 \cup S_2$. Therefore, the total Frank and Burgers vectors relative to the origin vanish, that is

$$\vec{\Omega} = \int_{S_1} \mathbf{q}^{(n)} dS + \int_{S_2} \mathbf{q}^{(n)} dS = 0 \tag{211}$$

$$\mathbf{B}_O = \int_{S_1} \left[ \mathbf{r} \times \mathbf{q}^{(n)} + \mathbf{a}^{(n)} \right] dS + \int_{S_2} \left[ \mathbf{r} \times \mathbf{q}^{(n)} + \mathbf{a}^{(n)} \right] dS = 0 \tag{212}$$

In terms of the Frank- and Burgers-stresses (disclination and dislocation densities), these relations become

$$\vec{\Omega} = \int_{S_1} d\mathbf{S} \bullet \mathbf{\theta} + \int_{S_2} d\mathbf{S} \bullet \mathbf{\theta} = 0 \tag{213}$$

$$\mathbf{B}_O = \int_{S_1} d\mathbf{S} \bullet \left[ \mathbf{\alpha} - \mathbf{\theta} \times \mathbf{r} \right] + \int_{S_2} d\mathbf{S} \bullet \left[ \mathbf{\alpha} - \mathbf{\theta} \times \mathbf{r} \right] = 0 \tag{214}$$

We notice these relations can also be written as

$$\vec{\Omega} = \vec{\Omega}^1 + \vec{\Omega}^2 = 0 \tag{215}$$

$$\mathbf{B}_O = \mathbf{B}_O^1 + \mathbf{B}_O^2 = 0 \tag{216}$$

where the total Frank true vectors and the total Burgers pseudo vectors on these surfaces are

$$\vec{\Omega}^1 = \int_{S_1} \mathbf{q}^{(n)} dS = \int_{S_1} d\mathbf{S} \bullet \mathbf{\theta} \qquad \mathbf{B}_O^1 = \int_{S_1} \left[ \mathbf{r} \times \mathbf{q}^{(n)} + \mathbf{a}^{(n)} \right] dS = \int_{S_1} d\mathbf{S} \bullet \left[ \mathbf{\alpha} - \mathbf{\theta} \times \mathbf{r} \right] \tag{217}$$

$$\vec{\Omega}^2 = \int_{S_2} \mathbf{q}^{(n)} dS = \int_{S_2} d\mathbf{S} \bullet \mathbf{\theta} \qquad \mathbf{B}_O^2 = \int_{S_2} \left[ \mathbf{r} \times \mathbf{q}^{(n)} + \mathbf{a}^{(n)} \right] dS = \int_{S_2} d\mathbf{S} \bullet \left[ \mathbf{\alpha} - \mathbf{\theta} \times \mathbf{r} \right] \tag{218}$$

It is obvious that the vectors $\vec{\Omega}^2 = -\vec{\Omega}^1$ and $\mathbf{B}_O^2 = -\mathbf{B}_O^1$ are not necessarily zero. Now if we assume the inner surface $S_2$ shrinks to some point, such that $\vec{\Omega}^2$ and $\mathbf{B}_O^2$ remain constant, they represent external concentrated Frank vector $\vec{\Omega}^2$ and Burgers vector $\mathbf{b}^2 = \mathbf{B}_O^2$ at that point, respectively. Therefore, it can be postulated that these concentrated external defects create elastic-plastic deformation in the body, which in turn results in internal continuous disclination density tensor $\mathbf{\theta}$ and dislocation density tensor $\mathbf{\alpha}$ (i.e., Frank- and Burgers-stress distributions) in the body.

### 7.2. Inconsistencies of the original continuous defect theory

In the original continuous defect theory (CDT), the disclination and dislocation density tensors (or the Frank- and Burgers-stress tensors) are

$$\mathbf{\theta} = -\nabla \times \mathbf{k}^P \qquad \theta_{ij} = -\varepsilon_{ikl} k^P_{lj,k} \qquad (219)$$

and

$$\mathbf{\alpha} = -\nabla \times \mathbf{e}^P + \mathbf{k}^{Pt} - tr(\mathbf{k}^P)\mathbf{I} \qquad \alpha_{ij} = -\varepsilon_{ikl} e^P_{lj,k} + k^P_{ji} - k^P_{ll}\delta_{ij} \qquad (220)$$

or

$$\mathbf{\alpha} = -\mathbf{\varepsilon} : (\nabla \mathbf{e}^P + \mathbf{k}^P \bullet \mathbf{\varepsilon}) \qquad \alpha_{ij} = -\varepsilon_{ikl}\left(e^P_{lj,k} + \varepsilon_{ljm} k^P_{km}\right) \qquad (221)$$

We have noticed that this continuous defect theory (CDT) is dual to the the indeterminate Mindlin-Tiersten-Koiter couple stress theory (MTK-CST). However, as mentioned previously, MTK-CST suffers from some serious inconsistencies, which are resolved in C-CST by establishing that the couple-stress pseudo (axial) tensor $\mathbf{\mu}$ is skew-symmetric. Is its dual quantity, the dislocation pseudo (axial) tensor $\mathbf{\alpha}$ (i.e., Burgers-stress tensor), also skew-symmetric? Here we first demonstrate that there are some fundamental inconsistencies and ambiguities in the orginal continuous defect theory (CDT) of Anthony (1970) and deWit (1970, 1973a-c), as presented above.

In the original continuous defect theory, the disclination density tensor $\mathbf{\theta}$ and dislocation density tensor $\mathbf{\alpha}$ each have nine components, as seen in (219) and (220). Therefore, the state of defect is described by 18 components altogether, but we have only six continuity equations (209) and (210). As demonstrated, the Burgers vector continuity equation gives the skew-symmetric part of the disclination density tensor $\mathbf{\Theta} = \mathbf{\theta}_{[\ ]}$ awkwardly in terms of nine components of the dislocation density tensor $\mathbf{\alpha}$, that is,

$$\mathbf{\Theta} = \mathbf{\theta}_{[\ ]} = \frac{1}{2}\mathbf{\varepsilon} \bullet (\nabla \bullet \mathbf{\alpha}) \qquad \Theta_{ij} = \theta_{[ij]} = -\frac{1}{2}\varepsilon_{ijk}\alpha_{lk,l} \qquad (222)$$

This relation actually indicates the interrelationship between the twist disclination vector $\vec{\Theta} = \frac{1}{2}\mathbf{\varepsilon} : \mathbf{\theta}$ and the dislocation density tensor $\mathbf{\alpha}$. However, we notice that this relation does not seem physically and mathematically consistent. The pseudo vector $\vec{\Theta}$ should be derived from the curl of a true vector or a second order skew-symmetric pseudo tensor. This clearly shows that the form of the dislocation density tensor (Burgers-stress tensor) $\mathbf{\alpha}$ should be skew-symmetric.

We also notice that a general form for the dislocation density (Burgers-stress) tensor $\mathbf{\alpha}$ in CDT cannot represent the internal defect properly. As explained, the defect at each point on a surface element $dS$ with unit normal vector $\mathbf{n}$ is specified by means of a Frank true vector $d\mathbf{\Omega} = \mathbf{q}^{(n)}dS$ and a Burgers pseudo vector $d\mathbf{b} = \mathbf{a}^{(n)}dS$. As a result, there are six types of continuous defects described by Frank and Burgers vectors on the surface element $dS$, which may be symbolically shown by the six discrete disclinations and dislocations in Figures 2 and 3. However, there is some issue with the normal component of the Burgers pseudo vector

$$d\mathbf{b} \bullet \mathbf{n} = \mathbf{a}^{(n)} \bullet \mathbf{n}dS = \mathbf{n} \bullet \mathbf{\alpha} \bullet \mathbf{n}dS \qquad db_i n_i = a_i^{(n)} n_i dS = \alpha_{ji} n_j n_i dS \qquad (223)$$

corresponding to the climbing edge dislocation, which can be explained as follows.

In crystal plasticity, the dislocations are considered as the result of the slip of crystal planes on each other. This means that the discrete gliding edge dislocation and screw dislocation (Figure 3a,c) can represent the infinitesimal continuous dislocation defect symbolically. However, the discrete climbing edge dislocation (Figure 3b), which requires violating the compatibility of the total deformation by breaking the body, does not have any role in crystal plasticity. This means the infinitesimal continuous climbing edge dislocation $d\mathbf{b} \bullet \mathbf{n}$ should not appear in a consistent continuous defect theory. Interestingly, some authors have been reluctant to consider the discrete climbing edge dislocation relevant in crystal plasticity, which clearly shows that there has been some disagreement on this type of dislocation in a continuous defect theory. For example, Lubliner (2008) and Berbenni et al. (2014) have considered only the gliding edge and screw dislocations to be relevant in a proper continuous defect theory, and thus in crystal plasticity.

This speculation suggests that in a consistent continuous defect theory, the climbing edge dislocation should not exist at any point on any arbitrary surface, that is

$$\mathbf{a}^{(n)} \bullet \mathbf{n} = \mathbf{n} \bullet \boldsymbol{\alpha} \bullet \mathbf{n} = 0, \qquad a_i^{(n)} n_i = \alpha_{ji} n_j n_i = 0 \qquad (224)$$

Since the tensor $n_j n_i$ is an arbitrary symmetric second order tensor of rank one, the relation (224) requires the Burgers-stress or dislocation tensor to be skew-symmetric

$$\boldsymbol{\alpha}^t = -\boldsymbol{\alpha} \qquad \alpha_{ji} = -\alpha_{ij} \qquad (225)$$

Interestingly, this result agrees with our previous speculation based on duality of geometry and statics in consistent couple stress theory. It also agrees with our suggestion that the vector $\vec{\Theta}$ should be derived from the curl of a vector or skew-symmetric dislocation tensor. As a result, it seems there can be only five types of defects in continuous defect theory: three disclinations and two dislocations. However, these results need to be established more rigorously. In the next section, we will prove these properties by using the consistent couple stress theory (C-CST) directly in continuous defect theory (CDT).

We also notice that there is some ambiguity concerning classical defect theory. Classical continuum mechanics can be recovered by neglecting the couple-stress tensor $\boldsymbol{\mu}$ in the indeterminate Mindlin-Tiersten-Koiter couple stress theory or in C-CST. Therefore, based on the duality of geometry and statics in Table 1, one can deduce that classical continuous defect theory is obtained by neglecting its dual quantity, the continuous dislocation density (Burgers-stress) tensor $\boldsymbol{\alpha}$, in the general continuous defect theory. Interestingly, this result shows that in classical defect theory, there can be only a symmetric disclination density (Frank-stress) tensor $\boldsymbol{\theta}$. However, this result has not been mentioned previously in the literature, which could be attributed to the lack of confidence in the original MTK-CST and the duality of geometry and statics of continuous defect theory.

Contrary to this result, the literature shows that classical defect theory has been taken as a theory based on the dislocation density tensor **α** (Nabarro, 1967; Kröner, 1970; deWit, 1973a). Interestingly, deWit (1973a) clearly considered a general continuous defect theory (CDT) with disclination and dislocation density tensors as an extension of a classical defect theory based on the dislocation density tensor. The matter has been so unsettled that even at the beginning of the 21st century, Kröner (2001) still presents a classical continuous defect theory of dislocations, while admitting that there are many shortcomings. The concept of dislocation is usually presented in manuscripts along with classical elasticity and plasticity. Therefore, no one has expected that there must be no continuous dislocation density (or Burgers-stress) in classical continuum mechanics and that any dislocation-based classical theory contradicts the duality between geometry and statics. The results in Sections 8 and 9 will show that this confusion can be attributed to inconsistencies in the original Mindlin-Tiersten-Koiter couple stress theory.

## 8. Consistent continuous defect theory

We have presented the original continuous defect theory (CDT) of Anthony (1970) and deWit (1970) from a new perspective, and investigated its inconsistencies. Let us summarize the elements of the original CDT as follows:

The compatibility conditions for the strain tensor **e** and the bend-twist tensor **k** are

$$\nabla \times \mathbf{k} = 0 \qquad \varepsilon_{ikl} k_{lj,k} = 0 \qquad (226)$$

$$\nabla \times \mathbf{e} + tr(\mathbf{k})\mathbf{I} - \mathbf{k}^t = 0 \qquad \varepsilon_{ikl} e_{lj,k} + k_{ll}\delta_{ij} - k_{ji} = 0 \qquad (227)$$

These are necessary and sufficient conditions for the existence of fields of displacement **u** and rotation $\vec{\omega}$ in a simply connected body. By using Weingarten's theorem, the disclination density tensor **θ** and dislocation density tensor **α** (i.e., Frank- and Burgers-stress tensors) have been defined as

$$\boldsymbol{\theta} = -\nabla \times \mathbf{k}^P \qquad \theta_{ij} = -\varepsilon_{ikl} k^P_{lj,k} \qquad (228)$$

$$\boldsymbol{\alpha} = -\nabla \times \mathbf{e}^P - tr(\mathbf{k}^P)\mathbf{I} + \mathbf{k}^{Pt} \qquad \alpha_{ij} = -\varepsilon_{ikl} e^P_{lj,k} - k^P_{ll}\delta_{ij} + k^P_{ji} \qquad (229)$$

The disclination density tensor $\boldsymbol{\theta}$ can be expressed as

$$\boldsymbol{\theta} = -\nabla \times \boldsymbol{\alpha}^t - \boldsymbol{\eta} + \boldsymbol{\varepsilon} \bullet \nabla\left[tr\left(\mathbf{k}^P\right)\right] \qquad \theta_{ij} = -\varepsilon_{ikl}\alpha_{jl,k} - \eta_{ij} + \varepsilon_{ijk}k^P_{ll,k} \qquad (230)$$

where the strain incompatibility tensor $\boldsymbol{\eta}$ is

$$\boldsymbol{\eta} = \nabla \times \mathbf{e}^E \times \nabla = -\nabla \times \mathbf{e}^P \times \nabla \qquad \eta_{ij} = -\varepsilon_{ikl}\varepsilon_{jmn}e^E_{ln,km} = \varepsilon_{ikl}\varepsilon_{jmn}e^P_{ln,km} \qquad (231)$$

The continuity equation for the incompatibility tensor $\boldsymbol{\eta}$ follows from its definition (231) as

$$\nabla \bullet \boldsymbol{\eta} = 0 \qquad \eta_{ji,j} = 0 \qquad (232)$$

It should be emphasized, contrary to Anthony (1970) and deWit (1970, 1973a-c), we have never decomposed the rotation vector $\vec{\omega}$ into elastic $\vec{\omega}^E$ and plastic $\vec{\omega}^P$ parts in this development, because $\vec{\omega}$ represents degrees of freedom.

The continuity condition for disclination and dislocation density tensors are

$$\nabla \bullet \boldsymbol{\theta} = 0 \qquad \theta_{ji,j} = 0 \qquad (233)$$

$$\nabla \bullet \boldsymbol{\alpha} + \boldsymbol{\varepsilon}:\boldsymbol{\theta} = 0 \qquad \alpha_{ji,j} + \varepsilon_{ijk}\theta_{jk} = 0 \qquad (234)$$

Meanwhile, in couple stress theory, the governing equilibrium equations are

$$\nabla \bullet \boldsymbol{\sigma} = 0 \qquad \sigma_{ji,j} = 0 \qquad (235)$$

$$\nabla \bullet \boldsymbol{\mu} + \boldsymbol{\varepsilon}:\boldsymbol{\sigma} = 0 \qquad \mu_{ji,j} + \varepsilon_{ijk}\sigma_{jk} = 0 \qquad (236)$$

which are clearly dual to (233) and (234). Table 1 summarizes the dualism between the geometry and statics of CDT in the original Mindlin-Tiersten-Koiter couple stress theory (deWit, 1970). Now by using consistent couple stress theory (C-CST), we develop the new consistent continuous defect theory C-CDT in a systematic manner.

Based on Weingarten's theorem, the plastic strain tensor $\mathbf{e}^P$ and plastic bend-twist tensor $\mathbf{k}^P$ are the fundamental kinematical quantities in defining the disclination density tensor $\boldsymbol{\theta}$ and dislocation density tensor $\boldsymbol{\alpha}$. As demonstrated, the dislocation density tensor $\boldsymbol{\alpha}$ represents the distortion incompatibility as

$$\boldsymbol{\alpha} = -\nabla \times \mathbf{e}^P + \mathbf{k}^{Pt} - tr(\mathbf{k}^P)\mathbf{I} \qquad \alpha_{ij} = -\varepsilon_{ikl} e^P_{lj,k} + k^P_{ji} - k^P_{ll}\delta_{ij} \qquad (237)$$

Are the kinematical quantities $\mathbf{e}^P$ and $\mathbf{k}^P$ independent of each other completely? In the inconsistent indeterminate original MTK-CST, the total strain tensor $\mathbf{e}$ and bend-twist tensor $\mathbf{k}$ are the fundamental measures of deformation. This can be seen from the principle of virtual work (116) for this formulation. By considering the elastic and plastic parts of the measures of deformation $\mathbf{e}$ and $\mathbf{k}$, this principle can be written as

$$\int_S \left[ \mathbf{t}^{(n)} \bullet \delta \mathbf{u} + \mathbf{m}^{(n)} \bullet \delta \vec{\omega} \right] dS = \int_V \left[ \boldsymbol{\sigma}:\delta\mathbf{e}^E + \boldsymbol{\mu}:\delta\mathbf{k}^E \right] dV + \int_V \left[ \boldsymbol{\sigma}:\delta\mathbf{e}^P + \boldsymbol{\mu}:\delta\mathbf{k}^P \right] dV \qquad (238)$$

This shows that in MTK-CST the strain tensor $\mathbf{e}^P$ and bend-twist tensor $\mathbf{k}^P$ are completely independent of each other, because these contribute to damage represented by their corresponding plastic work. This means although the relation

$$\mathbf{k} = tr(\mathbf{k})\mathbf{I} - \mathbf{e}\times\nabla \qquad k_{ij} = k_{ll}\delta_{ij} + \varepsilon_{jkl} e_{il,k} \qquad (239)$$

holds for the total bend-twist tensor, it does not hold separately for the elastic $\mathbf{k}^E$ and plastic $\mathbf{k}^P$ parts, as

$$\cancel{\mathbf{k}^E = tr(\mathbf{k}^E)\mathbf{I} - \mathbf{e}^E\times\nabla} \qquad \cancel{k^E_{ij} = k^E_{ll}\delta_{ij} + \varepsilon_{jkl} e^E_{il,k}} \qquad (240)$$

$$\cancel{\mathbf{k}^P = tr(\mathbf{k}^P)\mathbf{I} - \mathbf{e}^P\times\nabla} \qquad \cancel{k^E_{ij} = k^P_{ll}\delta_{ij} + \varepsilon_{jkl} e^P_{il,k}} \qquad (241)$$

Therefore, the individual bend-twist tensors $\mathbf{k}^E$ and $\mathbf{k}^P$ are incompatible in indeterminate MTK-CST. However, MTK-CST is an inconsistent theory, and the bend-twist tensor $\mathbf{k}$ is not the consistent bending measure of deformation.

Instead, within Consistent Couple Stress Theory (C-CST), the couple-stress tensor $\boldsymbol{\mu}$ is skew-symmetric, and the consistent measures of deformation are the strain tensor $\mathbf{e}$ and mean curvature tensor $\boldsymbol{\kappa}$. This can be seen from the principle of virtual work (118) for this formulation. By considering the elastic and plastic parts of the measures of deformation $\mathbf{e}$ and $\boldsymbol{\kappa}$, this principle can be written as

$$\int_S \left[ \mathbf{t}^{(n)} \bullet \delta \mathbf{u} + \mathbf{m}^{(n)} \bullet \delta \vec{\omega} \right] dS = \int_V \left[ \boldsymbol{\sigma} : \delta \mathbf{e}^E + \boldsymbol{\mu} : \delta \boldsymbol{\kappa}^E \right] dV + \int_V \left[ \boldsymbol{\sigma} : \delta \mathbf{e}^P + \boldsymbol{\mu} : \delta \boldsymbol{\kappa}^P \right] dV \qquad (242)$$

or

$$\int_S \left[ \mathbf{t}^{(n)} \bullet \delta \mathbf{u} + \mathbf{m}^{(n)} \bullet \delta \vec{\omega} \right] dS = \int_V \left[ \boldsymbol{\sigma} : \delta \mathbf{e}^E + 2\vec{\boldsymbol{\mu}} \bullet \delta \vec{\boldsymbol{\kappa}}^E \right] dV + \int_V \left[ \boldsymbol{\sigma} : \delta \mathbf{e}^P + 2\vec{\boldsymbol{\mu}} \bullet \delta \vec{\boldsymbol{\kappa}}^P \right] dV \qquad (243)$$

Since in C-CST, the plastic strain tensor $\mathbf{e}^P$ and plastic mean curvature tensor $\boldsymbol{\kappa}^P$ contribute to the damage specified by plastic work, these quantities are independent of each other. Consequently, although the relations

$$\boldsymbol{\kappa} = -\frac{1}{2}(\mathbf{e} \times \nabla + \nabla \times \mathbf{e}) \qquad \kappa_{ij} = \frac{1}{2}\left( \varepsilon_{jkl} e_{il,k} - \varepsilon_{ikl} e_{jl,k} \right) \qquad (244)$$

or

$$\vec{\boldsymbol{\kappa}} = \frac{1}{2}\left[ \nabla tr(\mathbf{e}) - \nabla \bullet \mathbf{e} \right] \qquad \kappa_i = \frac{1}{2}\left( e_{kk,i} - e_{ki,k} \right) \qquad (245)$$

hold for total quantities, these relations do not hold for elastic $\boldsymbol{\kappa}^E$ and plastic $\boldsymbol{\kappa}^P$ parts as

$$\cancel{\boldsymbol{\kappa}^E = -\frac{1}{2}\left( \mathbf{e}^E \times \nabla + \nabla \times \mathbf{e}^E \right)} \qquad \cancel{\kappa_{ij}^E = \frac{1}{2}\left( \varepsilon_{jkl} e_{il,k}^E - \varepsilon_{ikl} e_{jl,k}^E \right)} \qquad (246)$$

$$\cancel{\boldsymbol{\kappa}^P = -\frac{1}{2}\left( \mathbf{e}^P \times \nabla + \nabla \times \mathbf{e}^P \right)} \qquad \cancel{\kappa_{ij}^E = \frac{1}{2}\left( \varepsilon_{jkl} e_{il,k}^P - \varepsilon_{ikl} e_{jl,k}^P \right)} \qquad (247)$$

or

$$\cancel{\vec{\boldsymbol{\kappa}}^E = \frac{1}{2}\left[ \nabla tr(\mathbf{e}^E) - \nabla \bullet \mathbf{e}^E \right]} \qquad \cancel{\kappa_i^E = \frac{1}{2}\left( e_{kk,i}^E - e_{ki,k}^E \right)} \qquad (248)$$

$$\cancel{\vec{\boldsymbol{\kappa}}^P = \frac{1}{2}\left[ \nabla tr(\mathbf{e}^P) - \nabla \bullet \mathbf{e}^P \right]} \qquad \cancel{\kappa_i^P = \frac{1}{2}\left( e_{kk,i}^P - e_{ki,k}^P \right)} \qquad (249)$$

Therefore, the mean curvature tensors $\boldsymbol{\kappa}^E$ and $\boldsymbol{\kappa}^P$ or mean curvature vectors $\vec{\boldsymbol{\kappa}}^E$ and $\vec{\boldsymbol{\kappa}}^P$ are incompatible.

On the other hand, in C-CST, the symmetric torsion tensor $\chi$ is not a fundamental measure of deformation, because it does not contribute to the internal virtual work. This is despite the fact that this tensor can also be decomposed into elastic $\chi^E$ and plastic $\chi^P$ parts as

$$\chi = \chi^E + \chi^P \qquad \chi_{ij} = \chi_{ij}^E + \chi_{ij}^P \qquad (250)$$

Therefore, we notice that although the plastic torsion $\chi^P$ can exist in C-CST, it is not independent of the plastic strain through $\nabla \times \mathbf{e}^P$. This means both the elastic and plastic torsion tensors $\chi^E$ and $\chi^P$ are compatible. Hence, the relation

$$\chi = \frac{1}{2}(-\mathbf{e} \times \nabla + \nabla \times \mathbf{e}) \qquad \chi_{ij} = \frac{1}{2}\left(\varepsilon_{jkl} e_{il,k} + \varepsilon_{ikl} e_{jl,k}\right) \qquad (251)$$

holds not only for $\chi$, but also for its elastic and plastic parts $\chi^E$ and $\chi^P$, as

$$\chi^E = \frac{1}{2}\left(-\mathbf{e}^E \times \nabla + \nabla \times \mathbf{e}^E\right) \qquad \chi_{ij}^E = \frac{1}{2}\left(\varepsilon_{jkl} e_{il,k}^E + \varepsilon_{ikl} e_{jl,k}^E\right) \qquad (252)$$

$$\chi^P = \frac{1}{2}\left(-\mathbf{e}^P \times \nabla + \nabla \times \mathbf{e}^P\right) \qquad \chi_{ij}^P = \frac{1}{2}\left(\varepsilon_{jkl} e_{il,k}^P + \varepsilon_{ikl} e_{jl,k}^P\right) \qquad (253)$$

We notice that the relations (251)-(253) show that the total torsion tensor $\chi$, and its elastic and plastic parts $\chi^E$ and $\chi^P$ are deviatoric, that is

$$tr(\chi) = tr(\chi^E) = tr(\chi^P) = 0 \qquad \chi_{ii} = \chi_{ii}^E = \chi_{ii}^P = 0 \qquad (254)$$

Therefore, in a consistent continuous defect theory (C-CDT), only the mean curvature tensors $\kappa^E$ and $\kappa^P$, which represent the skew-symmetric parts of the bend-twist tensors $\mathbf{k}^E$ and $\mathbf{k}^P$, and the strain tensors $\mathbf{e}^E$ and $\mathbf{e}^P$ are incompatible. Meanwhile, in this new defect theory, the torsion tensors $\chi^E$ and $\chi^P$, which represent the symmetric parts of the bend-twist tensors $\mathbf{k}^E$ and $\mathbf{k}^P$, are compatible. Interestingly, as a result, we notice the bend-twist tensors $\mathbf{k}$, $\mathbf{k}^E$ and $\mathbf{k}^P$ are also deviatoric, that is

$$tr(\mathbf{k}) = tr(\mathbf{k}^E) = tr(\mathbf{k}^P) = 0 \qquad\qquad k_{ii} = k_{ii}^E = k_{ii}^P = 0 \qquad (255)$$

Table 2 summarizes the characteristics of the important kinematic fields in C-CST amd C-CDT.

**Table 2. Kinematic fields in consistent couple stress theory (C-CST) and consistent continuous defect theory (C-CDT)**

| Field | True or pseudo tensor | Degrees of freedom | Fundamental measure of deformation | Valid decomposition to elastic and plastic parts | Compatibility of elastic and plastic parts |
|---|---|---|---|---|---|
| Displacement vector $\mathbf{u}$ | True | Yes | No | No | - |
| Rotation tensor $\boldsymbol{\omega}$ | True | Yes | No | No | - |
| Rotation vector $\vec{\boldsymbol{\omega}}$ | Pseudo | Yes | No | No | - |
| Distortion tensor $\boldsymbol{\beta}$ | True | No | No | No | - |
| Strain tensor $\mathbf{e}$ | True | No | Yes | Yes | No |
| Bend-twist tensor $\mathbf{k}$ | Pseudo | No | No | Yes | No |
| Torsion tensor $\boldsymbol{\chi}$ | Pseudo | No | No | Yes | Yes |
| Mean curvature tensor $\boldsymbol{\kappa}$ | Pseudo | No | Yes | Yes | No |
| Mean curvature vector $\vec{\boldsymbol{\kappa}}$ | True | No | Yes | Yes | No |

It should be emphasized that in the original continuous defect theory (CDT) of Anthony (1970) and deWit (1970), the bend-twist tensors $\mathbf{k}^E$ and $\mathbf{k}^P$ are not necessarily deviatoric. This is the reason why $tr(\mathbf{k}^P)$ has appeared in relations (230) and (229) for the disclination density tensor $\boldsymbol{\theta}$ and dislocation density tensor $\boldsymbol{\alpha}$ in that theory, respectively.

Next, let us investigate the consequence of these facts on the character of the dislocation density tensor (Burgers-stress) $\boldsymbol{\alpha}$ and disclination density tensor (i.e., Frank-stress) $\boldsymbol{\theta}$ in C-CDT.

We notice that the dislocation density (Burgers-stress) tensor $\boldsymbol{\alpha}$ in (237) can also be written as

$$\boldsymbol{\alpha} = -\nabla \times \mathbf{e}^P + \boldsymbol{\kappa}^{Pt} + \boldsymbol{\chi}^P - tr(\mathbf{k}^P)\mathbf{I} \qquad \alpha_{ij} = -\varepsilon_{ikl} e^P_{jl,k} + \kappa^P_{ji} + \chi^P_{ij} - k^P_{ll}\delta_{ij} \qquad (256)$$

By using the compatibility relation (253) for $\boldsymbol{\chi}^P$ in the dislocation density relation (256) and noticing $tr(\mathbf{k}^P) = 0$ from (255), we obtain

$$\boldsymbol{\alpha} = -\nabla \times \mathbf{e}^P + \frac{1}{2}\left(-\mathbf{e}^P \times \nabla + \nabla \times \mathbf{e}^P\right) - \boldsymbol{\kappa}^P \qquad \alpha_{ij} = -\varepsilon_{ikl} e^P_{jl,k} + \frac{1}{2}\left(\varepsilon_{ikl} e^P_{jl,k} + \varepsilon_{jkl} e^P_{il,k}\right) - \kappa^P_{ij} \qquad (257)$$

After simplifying this relation, the dislocation pseudo tensor becomes

$$\boldsymbol{\alpha} = -\left[\boldsymbol{\kappa}^P + \frac{1}{2}\left(\nabla \times \mathbf{e}^P + \mathbf{e}^P \times \nabla\right)\right] \qquad \alpha_{ij} = -\left[\kappa^P_{ij} - \frac{1}{2}\left(\varepsilon_{jkl} e^P_{il,k} - \varepsilon_{ikl} e^P_{jl,k}\right)\right] \qquad (258)$$

This is an amazing result, which states that the dislocation (Burgers-stress) pseudo tensor $\boldsymbol{\alpha}$ is the negative of the incompatibility of the skew-symmetric mean curvature tensor $\boldsymbol{\kappa}^P$. Therefore, we have rigorously established that the dislocation pseudo tensor $\boldsymbol{\alpha}$ is skew-symmetric, that is

$$\boldsymbol{\alpha}^t = -\boldsymbol{\alpha} \qquad \alpha_{ji} = -\alpha_{ij} \qquad (259)$$

This is the fundamental subtle character of the dislocation density pseudo tensor $\boldsymbol{\alpha}$, which has not been recognized previously. Therefore, the dislocation pseudo tensor is specified by only three independent components. Thus, the dislocation density tensor now can be written as

$$[\alpha_{ij}] = \begin{bmatrix} 0 & \alpha_{12} & \alpha_{13} \\ -\alpha_{12} & 0 & \alpha_{23} \\ -\alpha_{13} & -\alpha_{23} & 0 \end{bmatrix} \qquad (260)$$

Since the pseudo (axial) dislocation tensor $\boldsymbol{\alpha}$ is skew-symmetrical, one can introduce its corresponding dual true (polar) dislocation vector $\vec{\alpha}$ as

$$\vec{\alpha} = \frac{1}{2}\boldsymbol{\varepsilon}:\boldsymbol{\alpha} \qquad \alpha_i = \frac{1}{2}\varepsilon_{ijk}\alpha_{jk} \qquad (261)$$

with the dual relation

$$\boldsymbol{\alpha} = \boldsymbol{\varepsilon}\bullet\vec{\alpha} \qquad \alpha_{ij} = \varepsilon_{ijk}\alpha_k \qquad (262)$$

Therefore, the components of the dislocation density (Burgers-stress) true vector are

$$\alpha_1 = \alpha_{23} \qquad \alpha_2 = -\alpha_{13} \qquad \alpha_3 = \alpha_{12} \qquad (263)$$

Interestingly, the dislocation true vector $\vec{\alpha}$ can also be expressed as

$$\vec{\alpha} = -\left\{\vec{\kappa}^P - \frac{1}{2}\left[\nabla tr(\mathbf{e}^P) - \nabla\bullet\mathbf{e}^P\right]\right\} \qquad \alpha_i = -\left[\kappa_i^P - \frac{1}{2}\left(e^P_{kk,i} - e^P_{ki,k}\right)\right] \qquad (264)$$

This relation shows that the dislocation true vector $\vec{\alpha}$ equals the negative of the incompatibility of the mean curvature true vector $\vec{\kappa}^P$, as expected from the dual tensorial relation (258).

Since $\boldsymbol{\alpha}$ is skew-symmetric and $tr(\mathbf{k}^P) = 0$, the relation (230) for the disclination density tensor (i.e., Frank-stress tensor) $\boldsymbol{\theta}$ in C-CDT becomes

$$\boldsymbol{\theta} = \nabla\times\boldsymbol{\alpha} - \boldsymbol{\eta} \qquad \theta_{ij} = \varepsilon_{ikl}\alpha_{lj,k} - \eta_{ij} \qquad (265)$$

with incompatibility tensor $\boldsymbol{\eta} = -\nabla\times\mathbf{e}^P\times\nabla$. By noticing that the curl of the skew-symmetric dislocation density tensor $\boldsymbol{\alpha}$ is

$$\nabla\times\boldsymbol{\alpha} = (\nabla\bullet\vec{\alpha})\mathbf{I} - \vec{\alpha}\nabla \qquad \varepsilon_{ikl}\alpha_{lj,k} = \alpha_{k,k}\delta_{ij} - \alpha_{i,j} \qquad (266)$$

the consistent form of the disclination density tensor $\boldsymbol{\theta}$ in C-CDT becomes

$$\boldsymbol{\theta} = -\vec{\alpha}\nabla + (\nabla\bullet\vec{\alpha})\mathbf{I} - \boldsymbol{\eta} \qquad \theta_{ij} = -\alpha_{i,j} + \alpha_{k,k}\delta_{ij} - \eta_{ij} \qquad (267)$$

For the symmetric and skew-symmetric parts of the consistent disclination density tensor, we have

$$\boldsymbol{\theta}_{()} = -\frac{1}{2}(\vec{\alpha}\nabla + \nabla\vec{\alpha}) + (\nabla\bullet\vec{\alpha})\mathbf{I} - \boldsymbol{\eta} \qquad \theta_{(ij)} = -\alpha_{(i,j)} + \alpha_{k,k}\delta_{ij} - \eta_{ij} \qquad (268)$$

$$\boldsymbol{\theta}_{[\,]} = -\frac{1}{2}(\vec{\boldsymbol{\alpha}}\nabla - \nabla\vec{\boldsymbol{\alpha}}) \qquad \theta_{[ij]} = -\alpha_{[i,j]} \qquad (269)$$

By using the expressions (264) and (231) for the dislocation density vector $\vec{\boldsymbol{\alpha}}$ and incompatibility tensor $\boldsymbol{\eta}$, respectively, the explicit relations for disclination density tensor $\boldsymbol{\theta}$ and its symmetric $\boldsymbol{\theta}_{(\,)}$ and skew-symmetric $\boldsymbol{\Theta} = \boldsymbol{\theta}_{[\,]}$ parts in terms of the incompatible plastic strain tensor $\mathbf{e}^P$ and incompatible mean curvature vector $\vec{\boldsymbol{\kappa}}^P$ are obtained as

$$\boldsymbol{\theta} = \vec{\boldsymbol{\kappa}}^P \nabla - (\nabla \bullet \vec{\boldsymbol{\kappa}}^P)\mathbf{I}$$
$$-\frac{1}{2}\left\{ \begin{array}{l} -2\nabla^2 \mathbf{e}^P - \nabla\left[tr(\mathbf{e}^P)\right]\nabla + 2\nabla(\mathbf{e}^P \bullet \nabla) + (\nabla \bullet \mathbf{e}^P)\nabla \\ +\nabla^2\left[tr(\mathbf{e}^P)\right]\mathbf{I} - \nabla \bullet (\nabla \bullet \mathbf{e}^P)\mathbf{I} \end{array} \right\} \qquad (270)$$

$$\boldsymbol{\theta}_{(\,)} = \frac{1}{2}(\vec{\boldsymbol{\kappa}}^P \nabla + \nabla\vec{\boldsymbol{\kappa}}^P) - (\nabla \bullet \vec{\boldsymbol{\kappa}}^P)\mathbf{I}$$
$$-\frac{1}{2}\left\{ \begin{array}{l} -2\nabla^2 \mathbf{e}^P - \nabla\left[tr(\mathbf{e}^P)\right]\nabla + \frac{3}{2}\left[\nabla(\mathbf{e}^P \bullet \nabla) + (\nabla \bullet \mathbf{e}^P)\nabla\right] \\ \nabla^2\left[tr(\mathbf{e}^P)\right]\mathbf{I} - \nabla \bullet (\nabla \bullet \mathbf{e}^P)\mathbf{I} \end{array} \right\} \qquad (271)$$

$$\boldsymbol{\theta}_{[\,]} = \boldsymbol{\Theta} = \frac{1}{2}(\vec{\boldsymbol{\kappa}}^P \nabla - \nabla\vec{\boldsymbol{\kappa}}^P) - \frac{1}{4}\left[\nabla(\nabla \bullet \mathbf{e}^P) - (\nabla \bullet \mathbf{e}^P)\nabla\right] \qquad (272)$$

In terms of components, these relations become

$$\theta_{ij} = \kappa^P_{i,j} - \kappa^P_{k,k}\delta_{ij} - \frac{1}{2}\left[-2e^P_{ij,kk} - e^P_{kk,ij} + 2e^P_{jk,ik} + e^P_{ik,jk} + \delta_{ij}\left(e^P_{kk,ll} - e^P_{kl,kl}\right)\right] \qquad (273)$$

$$\theta_{(ij)} = \frac{1}{2}\left(\kappa^P_{i,j} + \kappa^P_{j,i}\right) - \kappa^P_{k,k}\delta_{ij} - \frac{1}{2}\left[-2e^P_{ij,kk} - e^P_{kk,ij} + \frac{3}{2}\left(e^P_{jk,ik} + e^P_{ik,jk}\right) + \delta_{ij}\left(e^P_{kk,ll} - e^P_{kl,kl}\right)\right] \qquad (274)$$

$$\theta_{[ij]} = \Theta_{ij} = \frac{1}{2}\left(\kappa^P_{i,j} - \kappa^P_{j,i}\right) - \frac{1}{4}\left(e^P_{jk,ik} - e^P_{ik,jk}\right) \qquad (275)$$

We also notice the relation for the trace of the disclination density or Frank-stress tensor as

$$tr(\boldsymbol{\theta}) = -2\nabla \bullet \vec{\boldsymbol{\kappa}}^P \qquad \theta_{ii} = -2\kappa^P_{i,i} \qquad (276)$$

Interestingly, for the twist disclination pseudo (axial) vector $\vec{\boldsymbol{\Theta}}$, we have

$$\vec{\Theta} = \frac{1}{2}\nabla \times \vec{\alpha} \qquad \Theta_i = \frac{1}{2}\varepsilon_{ijk}\alpha_{k,j} \qquad (277)$$

This relation shows that the twist disclination pseudo vector $\vec{\Theta}$ is the curl of the dislocation density true vector $\vec{\alpha}$, which agrees with our initial speculation that $\vec{\Theta}$ should be curl of a true vector.

We notice that in C-CDT, the Burgers-traction pseudo vector can be written as

$$\mathbf{a}^{(n)} = \vec{\alpha} \times \mathbf{n} \qquad a_i^{(n)} = \varepsilon_{ikj}\alpha_k n_j \qquad (278)$$

which shows that $\mathbf{a}^{(n)}$ is tangent to the surface. Therefore, the Burgers couple with Burgers pseudo vector

$$d\mathbf{b} = \mathbf{a}^{(n)} dS \qquad db_i = a_i^{(n)} dS \qquad (279)$$

has a slipping character. This remarkable result shows that the Burgers-traction pseudo vector $\mathbf{a}^{(n)}$ creates only sliding dislocations. This means that there are only two types of continuous dislocations, which correspond symbolically to the discrete gliding edge dislocation and screw dislocation in Figure 3a,c, respectively.

Therefore, the defect on a surface element $dS$ with unit normal vector $\mathbf{n}$ is specified by the infinitesimal Frank vector $d\vec{\Omega}$ and infinitesimal Burgers vector $d\mathbf{b}$, where

$$d\vec{\Omega} = \mathbf{q}^{(n)} dS = d\mathbf{S} \bullet \mathbf{\theta} \qquad d\Omega_i = q_i^{(n)} dS = \theta_{ji} n_j dS \qquad (280)$$

$$d\mathbf{b} = \mathbf{a}^{(n)} dS = \vec{\alpha} \times d\mathbf{S} \qquad db_i = a_i^{(n)} dS = \varepsilon_{ikj}\alpha_k n_j dS \qquad (281)$$

C-CST has provided the fundamental physical and mathematical reasons for the skew-symmetric character of the dislocation density (or Burgers-stress) pseudo tensor, after half a century of confusion. In fact, proper dislocation density is a true vector representing internal plastic induced sliding defects. We have rigorously established that there are only five types of defects in consistent continuous defect theory (C-CDT): three disclinations and two dislocations. Interestingly, Kleman and Lavrentovich, (2003) and Kleman and Friedel (2008) have speculated that there are five relevant types of defects in continuous defect theory: two disclinations and three dislocations. However, this choice of relevant types of defects is inconsistent with the theoretical development of C-CDT.

It is quite astonishing to recognize that Weingarten's theorem, along with C-CST, has led to the formulation of the theoretical construct of C-CDT for disclinations and dislocations, which is dual to the statics of C-CST. Table 3 summarizes the dualism between the geometry and statics of consistent continuous defect theory (C-CDT) in consistent couple stress theory (C-CST).

**Table 3. Dualism between geometry and statics of consistent continuous defect theory**

| Geometry<br>C-CDT | Statics<br>C-CST |
|---|---|
| Frank vector $\vec{\Omega}$ | Force vector $\mathbf{F}$ |
| Burgers vector $\mathbf{B}$ | Moment vector $\mathbf{M}$ |
| Frank-traction vector $\mathbf{q}^{(n)}$ | Force-traction vector $\mathbf{t}^{(n)}$ |
| Burgers-traction vector $\mathbf{a}^{(n)}$<br>$\mathbf{a}^{(n)} \bullet \mathbf{n} = 0$ | Couple-traction vector $\mathbf{m}^{(n)}$<br>$\mathbf{m}^{(n)} \bullet \mathbf{n} = 0$ |
| Disclination density tensor $\boldsymbol{\theta}$<br>(Frank-stress tensor) | Force-stress tensor $\boldsymbol{\sigma}$ |
| Dislocation density tensor $\boldsymbol{\alpha}$<br>(Burgers-stress tensor)<br>$\boldsymbol{\alpha}^t = -\boldsymbol{\alpha}$ | Couple-stress tensor $\boldsymbol{\mu}$<br><br>$\boldsymbol{\mu}^t = -\boldsymbol{\mu}$ |
| Governing equations<br>(Continuity equations)<br>$\nabla \bullet \boldsymbol{\theta} = 0$<br>$\nabla \bullet \boldsymbol{\alpha} + \boldsymbol{\varepsilon} : \boldsymbol{\theta} = 0$ | Governing equations<br>(Equilibrium equations)<br>$\nabla \bullet \boldsymbol{\sigma} = 0$<br>$\nabla \bullet \boldsymbol{\mu} + \boldsymbol{\varepsilon} : \boldsymbol{\sigma} = 0$ |
| $\vec{\Theta} = \frac{1}{2} \nabla \times \vec{\alpha}$ | $\mathbf{s} = \frac{1}{2} \nabla \times \vec{\mu}$ |
| $\boldsymbol{\theta} = \nabla \times \boldsymbol{\alpha} - \boldsymbol{\eta}$<br>$\boldsymbol{\alpha} = -\left[ \boldsymbol{\kappa}^P + \frac{1}{2}\left( \nabla \times \mathbf{e}^P + \mathbf{e}^P \times \nabla \right) \right]$<br>$\boldsymbol{\eta} = -\nabla \times \mathbf{e}^P \times \nabla$ | |

Recall that Weingarten's theorem shows that there are six types of discrete or external concentrated defects in a deformable body, represented by a Frank true vector $\vec{\Omega}$ and a Burgers pseudo vector $\mathbf{B}$. This is dual (analogous) to six types of independent external loading represented by concentrated force true vector $\mathbf{F}$ and couple with moment pseudo vector $\mathbf{M}$ in a rigid body with six degrees of freedom.

We have demonstrated, however, that there are only five independent types of continuous surface defects on a surface element $dS$ with unit normal vector $\mathbf{n}$ in a deformable body represented by the infinitesimal Frank true-vector $d\vec{\Omega} = \mathbf{q}^{(n)} dS = d\mathbf{S} \bullet \mathbf{\theta}$ and the Burgers couple with Burgers pseudo vector $d\mathbf{b} = \mathbf{a}^{(n)} dS = \vec{\alpha} \times d\mathbf{S}$. This is dual (analogous) to the five types of independent loadings on the surface element $dS$ represented by the infinitesimal force true vector $d\mathbf{F} = \mathbf{t}^{(n)} dS = d\mathbf{S} \bullet \mathbf{\sigma}$ and the infinitesimal couple pseudo vector $d\mathbf{M} = \mathbf{m}^{(n)} dS = \vec{\mu} \times d\mathbf{S}$ corresponding to five independent degrees of freedom on the surface.

It should be emphasized that the continuous defect theory (CDT) of Anthony (1970) and deWit (1970) stands as a fundamental contribution in the development of C-CDT. This is obvious from the fact that elements of C-CDT are based on their original formulation of CDT. However, there have been some inconsistencies in the original CDT, which has created difficulty in the progress of continuous defect theory. These include taking the bend-twist tensor $\mathbf{k}$ as a fundamental measure of deformation, and decomposing the displacement and rotation vectors $\mathbf{u}$ and $\mathbf{\omega}$ into elastic and plastic parts. We notice that these troubles are analogous to the inconsistencies identified with the original MTK-CST. It has been very unfortunate that Mindlin, Tiersten and Koiter missed the skew-symmetric character of the couple-stress tensor, although they correctly stated the character of the boundary conditions in couple stress theory. In the next section, we investigate the character of disclination and dislocation density tensors (i.e., Frank- and Burgers- stress tensors) in the restricted consistent classical continuous defect theory, which is fully compatible with classical continuum mechanics.

## 9. Consistent classical continuous defect theory

As mentioned previously, classical continuum mechanics can be considered as a special case of the general C-CST, in which the couple-stress tensor **μ** is neglected. Therefore, based on the duality of geometry and statics in Table 3, it may be speculated that Consistent Classical Continuous Defect Theory (CC-CDT) is obtained by neglecting the continuous dislocation density tensor (Burgers-stress tensor) **α** in the general C-CDT presented above in Section 8. However, we establish this speculation rigorously by investigating the character of CC-CDT in classical continuum mechanics without using the duality of geometry and statics directly.

Since there are no couple-stresses in classical continuum theory, that is, $\boldsymbol{\mu} = 0$, the moment balance governing equation (236) results in the symmetric character of the force-stress tensor **σ** in classical theory. Thus,

$$\boldsymbol{\sigma}^t = \boldsymbol{\sigma} \qquad \sigma_{ji} = \sigma_{ij} \tag{282}$$

Therefore, the governing equation in classical theory is the force balance equation

$$\nabla \bullet \boldsymbol{\sigma} = 0 \qquad \sigma_{ji,j} = 0 \tag{283}$$

Since there is no couple-traction, i.e., $\mathbf{m}^{(n)} = 0$, the principle of virtual work (118) for classical continuum theory reduces to

$$\int_S \mathbf{t}^{(n)} \bullet \delta \mathbf{u} dS = \int_V \boldsymbol{\sigma} : \delta \mathbf{e} dV \tag{284}$$

This clearly shows that in classical continuum theory the mean curvature tensor **κ** is not a fundamental measure of deformation, because its energy conjugate couple-stress tensor **μ** does not exist. By considering the elastic and plastic parts of the measures of deformation **e**, this principle can be written as

$$\int_S \mathbf{t}^{(n)} \bullet \delta \mathbf{u} dS = \int_V \boldsymbol{\sigma} : \delta \mathbf{e}^E dV + \int_V \boldsymbol{\sigma} : \delta \mathbf{e}^P dV \tag{285}$$

We notice that although the elastic and plastic parts $\boldsymbol{\kappa}^E$ and $\boldsymbol{\kappa}^P$ can exist, these are compatible. Therefore, not only the plastic torsion tensor $\boldsymbol{\chi}^P$ is compatible, the plastic mean curvature tensor $\boldsymbol{\kappa}^P$ also is also compatible. However, the dislocation density (Burgers-stress) pseudo tensor $\boldsymbol{\alpha}$ is the negative of the incompatibility of the mean curvature tensor $\boldsymbol{\kappa}^P$. Therefore, there can be no continuous dislocation density tensor $\boldsymbol{\alpha}$ in classical continuous defect theory, that is,

$$\boldsymbol{\alpha} = -\left[\boldsymbol{\kappa}^P + \frac{1}{2}\left(\nabla \times \mathbf{e}^P + \mathbf{e}^P \times \nabla\right)\right] = 0 \qquad \alpha_{ij} = -\left[\kappa^P_{ij} - \frac{1}{2}\left(\varepsilon_{jkl} e^P_{il,k} - \varepsilon_{ikl} e^P_{jl,k}\right)\right] = 0 \qquad (286)$$

or

$$\vec{\alpha} = -\left\{\vec{\kappa}^P - \frac{1}{2}\left[\nabla tr\left(\mathbf{e}^P\right) - \nabla \bullet \mathbf{e}^P\right]\right\} = 0 \qquad \alpha_i = -\left[\kappa^P_i - \frac{1}{2}\left(e^P_{kk,i} - e^P_{ki,k}\right)\right] = 0 \qquad (287)$$

This result also shows that in CC-CDT, the elastic and plastic bend-twist tensors $\mathbf{k}^E$ and $\mathbf{k}^P$ are compatible, that is,

$$\mathbf{k}^E = -\mathbf{e}^E \times \nabla \qquad k^E_{ij} = \varepsilon_{jkl} e^E_{il,k} \qquad (288)$$

$$\mathbf{k}^P = -\mathbf{e}^P \times \nabla \qquad k^P_{ij} = \varepsilon_{jkl} e^P_{il,k} \qquad (289)$$

Consequently, the disclination density tensor (Frank-stress tensor) $\boldsymbol{\theta}$ from (228) reduces to

$$\begin{aligned}\boldsymbol{\theta} &= -\nabla \times \mathbf{k}^P \\ &= \nabla \times \mathbf{e}^P \times \nabla \\ &= -\boldsymbol{\eta}\end{aligned} \qquad (290)$$

Therefore, in consistent classical continuous defect theory (CC-CDT), the disclination density tensor (Frank-stress tensor) $\boldsymbol{\theta}$ is symmetric and equals the negative of the strain incompatibility tensor $\boldsymbol{\eta}$. This tensor can also be represented as

$$\begin{aligned}\theta_{ij} = -\eta_{ij} &= -\varepsilon_{ikl}\varepsilon_{jmn} e^P_{ln,km} \\ &= e^P_{ij,kk} + e^P_{kk,ij} - e^P_{jk,ik} - e^P_{ik,jk} - \delta_{ij}\left(e^P_{kk,ll} - e^P_{kl,kl}\right)\end{aligned} \qquad (291)$$

Interestingly, we notice

$$\theta_{ii} = e^P_{ik,ik} - e^P_{ii,kk} \qquad (292)$$

$$\theta_{ij} - \theta_{kk}\delta_{ij} = \left(-e^P_{ik,j} - e^P_{jk,i} + e^P_{ij,k}\right)_{,k} + e^P_{kk,ij} \qquad (293)$$

From the development above, there can be no Burgers-stress or dislocation density $\boldsymbol{\alpha}$ in classical continuous defect theory. This is truly a remarkable result. Furthermore, we have established this result without using the duality of geometry and statics in Table 3. In this classical version of consistent continuous defect theory, there is only the disclination density tensor $\boldsymbol{\theta}$, which now is symmetric. Thus,

$$\boldsymbol{\theta}^t = \boldsymbol{\theta} \qquad \theta_{ij} = \theta_{ji} \tag{294}$$

which satisfies

$$\nabla \bullet \boldsymbol{\theta} = 0 \qquad \theta_{ji,j} = 0 \tag{295}$$

We have demonstrated that classical continuum mechanics with symmetric force-stress tensor $\boldsymbol{\sigma}$ necessitates a classical continuous defect theory with symmetric Frank-stress or disclination density tensor $\boldsymbol{\theta}$. This means there is no internal plastic induced continuous dislocation density (Burgers-stress) tensor $\boldsymbol{\alpha}$ in the body, when there are no couple-stresses. Therefore, CC-CDT is integrable in the rotation, but not in the displacement. In this theory, there is only the Frank-traction vector $\mathbf{q}^{(n)}$, but there is no Burgers-traction vector, that is,

$$\mathbf{q}^{(n)} = \mathbf{n} \bullet \boldsymbol{\theta} \qquad q_i^{(n)} = \theta_{ji} n_j \tag{296}$$

$$\mathbf{a}^{(n)} = \mathbf{n} \bullet \boldsymbol{\alpha} = 0 \qquad a_i^{(n)} = \alpha_{ji} n_j = 0 \tag{297}$$

This means the continuous defects are the result of local internal inclination of planes relative to each other, not the result of slipping of planes. In CC-CDT, the defect on a surface element $dS$ with unit normal vector $\mathbf{n}$ is specified entirely by the infinitesimal Frank true vector

$$d\vec{\Omega} = \mathbf{q}^{(n)} dS = d\mathbf{S} \bullet \boldsymbol{\theta} \qquad d\Omega_i = q_i^{(n)} dS = \theta_{ji} n_j dS \tag{298}$$

while the infinitesimal Burgers vector is zero, that is,

$$d\mathbf{b} = 0 \qquad db_i = 0 \tag{299}$$

For a surface $S$, which is bounded by the closed curve $C$, the total Frank true vector $\vec{\Omega}$ and the Burgers pseudo vector $\mathbf{B}_O$ relative to the origin are given as

$$\vec{\Omega} = \int_S \mathbf{q}^{(n)} dS \qquad \Omega_i = \int_S q_i^{(n)} dS \tag{300}$$

$$\mathbf{B}_O = \int_S \mathbf{r} \times \mathbf{q}^{(n)} dS \qquad B_{Oi} = \int_S \varepsilon_{ijl} x_j q_l^{(n)} dS \qquad (301)$$

By using the relations for the Frank-traction vector in terms of the disclination density or Frank-stress tensor $\boldsymbol{\theta}$, these relations become

$$\vec{\boldsymbol{\Omega}} = \int_S d\mathbf{S} \bullet \boldsymbol{\theta} \qquad \Omega_i = \int_S \theta_{ji} n_j dS \qquad (302)$$

$$\mathbf{B}_O = \int_S d\mathbf{S} \bullet [-\boldsymbol{\theta} \times \mathbf{r}] \qquad B_{Oi} = \int_S \varepsilon_{ijl} x_j \theta_{kl} n_k dS \qquad (303)$$

For an arbitrary part of the material continuum occupying a volume $V$ enclosed by boundary surface $S$, the circuit $C$ shrinks to a point. Therefore, the total Frank true vector and the Burgers pseudo vector relative to any point vanish, that is, $\vec{\boldsymbol{\Omega}} = 0$ and $\mathbf{B} = 0$. In this case, the total Frank and Burgers vectors for this part of the body are written, respectively, as

$$\int_S \mathbf{q}^{(n)} dS = 0 \qquad \int_S q_i^{(n)} dS = 0 \qquad (304)$$

$$\int_S \mathbf{r} \times \mathbf{q}^{(n)} dS = 0 \qquad \int_S \varepsilon_{ijk} x_j q_k^{(n)} dS = 0 \qquad (305)$$

In terms of the disclination density tensor, these become

$$\int_S d\mathbf{S} \bullet \boldsymbol{\theta} = 0 \qquad \int_S \theta_{ji} n_j dS = 0 \qquad (306)$$

$$\int_S d\mathbf{S} \bullet [-\boldsymbol{\theta} \times \mathbf{r}] = 0 \qquad \int_S \varepsilon_{ijl} x_j \theta_{kl} n_k dS = 0 \qquad (307)$$

which result in the governing continuity equations

$$\nabla \bullet \boldsymbol{\theta} = 0 \qquad \theta_{ji,j} = 0 \qquad (308)$$

$$\boldsymbol{\varepsilon} : \boldsymbol{\theta} = 0 \qquad \varepsilon_{ijk} \theta_{jk} = 0 \qquad (309)$$

We notice that the Burgers continuity equation (309), establishes the symmetric character of the Frank-stress or disclination density tensor (294) within consistent classical continuous defect theory. From (290), we can recognize that the strain incompatibility tensor $\boldsymbol{\eta}$ satisfies the same set of continuity equations.

In this CC-CDT, the total Frank vector $\vec{\Omega}$ and total Burgers vector $\mathbf{B}_O$ on a surface $S$ bounded by the circuit $C$ are still defined as

$$\vec{\Omega} = -\oint_C d\mathbf{r} \bullet \mathbf{k}^P \qquad \Omega_i = -\oint_C k_{ji}^P dx_j \qquad (310)$$

$$\mathbf{B}_O = -\oint_C d\mathbf{r} \bullet \left[ \mathbf{e}^P - \mathbf{k}^P \times \mathbf{r} \right] \qquad B_{Oi} = -\oint_C \left( e_{ki}^P - \varepsilon_{ilj} k_{kl}^P x_j \right) dx_k \qquad (311)$$

Since the elastic and plastic bend-twist tensors $\mathbf{k}^E$ and $\mathbf{k}^P$ are compatible, the total Frank vector $\vec{\Omega}$ and total Burgers vector $\mathbf{B}_O$ can be expressed as

$$\vec{\Omega} = \oint_C d\mathbf{r} \bullet \left( \mathbf{e}^P \times \nabla \right) \qquad (312)$$

$$\mathbf{B}_O = -\oint_C d\mathbf{r} \bullet \left[ \mathbf{e}^P + \left( \mathbf{e}^P \times \nabla \right) \times \mathbf{r} \right] \qquad (313)$$

Table 4 summarizes the dualism between the geometry and statics of the proper version of classical continuous defect theory in classical continuum mechanics.

As mentioned previously, the same results for CC-CDT may be obtained from the inconsistent original continuous defect theory (CDT), where the dislocation density tensor $\boldsymbol{\alpha}$ is in general form. By neglecting the couple-stress tensor $\boldsymbol{\mu}$, the principle of virtual work (116) reduces to its classical counterpart (284) in classical continuum mechanics, which shows that the bend-twist tensor $\mathbf{k}$ is not a fundamental measure of deformation. By considering the elastic and plastic parts of the measures of deformation $\mathbf{e}$, this principle can be written as (285), which shows that although the elastic and plastic parts $\mathbf{k}^E$ and $\mathbf{k}^P$ can exist, these tensors are compatible. As a consequence, we obtain the same result that the continuous dislocation density tensor $\boldsymbol{\alpha}$ vanishes in classical continuous defect theory, that is,

$$\boldsymbol{\alpha} = -\nabla \times \mathbf{e}^P + \mathbf{k}^{Pt} - tr(\mathbf{k}^P)\mathbf{I} = 0 \qquad \alpha_{ij} = -\varepsilon_{ikl} e_{lj,k}^P + k_{ji}^P - k_{ll}^P \delta_{ij} = 0 \qquad (314)$$

**Table 4. Dualism between geometry and statics of consistent classical continuous defect theory**

| Geometry<br>CC-CDT | Statics<br>Classical Continuum Theory |
|---|---|
| Frank vector $\vec{\Omega}$ | Force vector $\mathbf{F}$ |
| Burgers vector $\mathbf{B}$ | Moment vector $\mathbf{M}$ |
| Frank-traction vector $\mathbf{q}^{(n)}$ | Force-traction vector $\mathbf{t}^{(n)}$ |
| Burgers-traction vector $\mathbf{a}^{(n)} = 0$ | Couple-traction vector $\mathbf{m}^{(n)} = 0$ |
| Disclination density tensor $\boldsymbol{\theta}$<br>(Frank-stress tensor) | Force-stress tensor $\boldsymbol{\sigma}$ |
| Dislocation density tensor $\boldsymbol{\alpha} = 0$<br>(Burgers-stress tensor) | Couple-stress tensor $\boldsymbol{\mu} = 0$ |
| Governing equations<br>(Continuity equations)<br>$\nabla \bullet \boldsymbol{\theta} = 0$<br>$\boldsymbol{\theta}^t = \boldsymbol{\theta}$ | Governing equations<br>(Equilibrium equations)<br>$\nabla \bullet \boldsymbol{\sigma} = 0$<br>$\boldsymbol{\sigma}^t = \boldsymbol{\sigma}$ |
| $\boldsymbol{\theta} = -\boldsymbol{\eta}$<br>$= -\nabla \times \mathbf{e}^P \times \nabla$ | |

Contrary to this conclusion, as mentioned above, classical defect theory has been taken historically as a theory based on the dislocation density (or Burgers-stress) tensor (Nabarro, 1967; Kröner, 1970; deWit, 1973a; Kroner, 2001).

It also should be emphasized that there is no problem with the existence of external concentrated or distributed dislocations within classical continuum mechanics. Even external climbing edge dislocation distributions can be used to model crack and contact problems within classical continuum mechanics (e.g., Moore and Hills, 2018). However, as demonstrated above, there can be no internal plastic induced distribution of dislocation density in classical continuum mechanics.

## 10. Conclusions

The recent development of consistent couple stress theory (C-CST) has provided the fundamental basis to resolve the existing troubles and confusions in continuous defect theory (CDT). By using some elements of the original CDT, we have been able to develop consistent continuous defect theory (C-CDT) with amazing consequences. C-CST and Weingarten's theorem play the central roles in defining the consistent disclination and dislocation density (Frank- and Burgers-stress) tensors in C-CDT. It turns out that in C-CDT, the dislocation density (or Burgers-stress) tensor is skew-symmetric and can be represented by a true (polar) vector. The fundamental step in this discovery is recognizing the fact that only the skew-symmetrical part of the bend-twist tensor is a measure of deformation, energy conjugate to the skew-symmetric couple-stress tensor. Although the full bend-twist tensor is an important kinematical quantity in Weingarten's theorem, it is not a fundamental measure of deformation.

Remarkably, the development presented here rigorously establishes the dualism between geometry and statics of C-CDT based on C-CST. Therefore, C-CDT may provide a fundamental basis to study multi-scale crystal plasticity from a continuum mechanics perspective. We notice that the size-effect in this continuum plasticity is interrelated directly with the dislocation density, not the disclination density tensor.

We have also demonstrated that a restriction to classical continuum mechanics with symmetric force-stress tensor would require a consistent classical continuous defect theory (CC-CDT) with symmetric disclination density tensor (or Frank-stress tensor). This means the internal continuous dislocation density tensor is zero throughout the entire body, when there are no couple-stresses. Therefore, this classical defect theory is integrable in the rotation, but not integrable in the displacement. We notice then that existence of a continuous dislocation (or Burgers-stress) tensor requires the existence of the couple-stress tensor. Interestingly, this remarkable result is further evidence for the reality of couple-stresses. Therefore, we have shown that contrary to the common belief, general consistent continuous defect theory (C-CDT) must be an extension of a classical continuous disclination defect theory. This is completely compatible with the dualism between C-CDT and the statics of C-CST.

Additional aspects of this new C-CDT, including constitutive relations and the motion of defects will be addressed in a forthcoming work. In a future paper, the geometrical character of the force- and couple-stress tensors, as continuous defects (Frank- and Burgers-stress or disclination and dislocation density tensors of a geometrical displacement-like field) will be investigated by using modified Günther stress function tensors, based on C-CDT. In this development, concentrated forces and concentrated couples are analogous to concentrated Frank and Burgers vectors of the corresponding geometrical field. In all of this development, it is very important to realize that consistent couple stress theory (C-CST) serves as a Rosetta Stone. By using this theory, we have been able to decipher some existing inconsistencies in continuous defect theory, and as a result develop fully consistent continuous defect theory (C-CDT).

## References

Anthony, K., Essmann, U., Seeger, A. and Träuble, H., 1968. Disclinations and the Cosserat-continuum with incompatible rotations. In Mechanics of generalized continua, 355-358. Springer, Berlin, Heidelberg.

Anthony, K.H., 1970. Die theorie der disklinationen. Archive for Rational Mechanics and Analysis, 39(1), 43-88.

Berbenni, S., Taupin, V., Djaka, K.S. and Fressengeas, C., 2014. A numerical spectral approach for solving elasto-static field dislocation and g-disclination mechanics. I Int. J. Solids Struct. 51(23-24), 4157-4175.

Cordier, P., Demouchy, S., Beausir, B., Taupin, V., Barou, F. and Fressengeas, C., 2014. Disclinations provide the missing mechanism for deforming olivine-rich rocks in the mantle. Nature, 507(7490), 51-56.

Cosserat, E., Cosserat, F., 1909. Théorie des Corps déformables. Hermann, Paris.

Eringen, A. C., 1968. Theory of micropolar elasticity, Fracture, vol 2, ed. H. Liebowitz, Academic Press, New York, 662-729.

deWit, R., 1960. The continuum theory of stationary dislocations. Solid State Phys. 10, 269–292.

deWit, R., 1970. Linear theory of static disclinations. J. Res. Nat. Bur. Stand. A. Phys. Chem. 77A (1), 651–680.

deWit, R., 1973. Theory of disclinations: II Continuous and discrete disclinations in anisotropic elasticity. J. Res. Nat. Bur. Stand. A. Phys. Chem. 77A (1), 49–100.

deWit, R., 1973. Theory of disclinations: III Continuous and discrete disclinations in isotropic elasticity. J. Res. Nat. Bur. Stand. A. Phys. Chem. 77A (3), 359–368.

deWit, R., 1973. Theory of disclinations: IV Straight disclinations. J. Res. Nat. Bur. Stand. A. Phys. Chem. 77A (5), 607–658.

Frank, F.C., 1958. I. Liquid crystals. On the theory of liquid crystals. Discuss. Faraday Soc., 25, 19-28.

Günther, W., 1958. Zur Statik und Kinematik des Cosseratschen Kontinuums. Abh. Braunschweig. Wiss. Ges. 10, 195–213.

Hadjesfandiari, A. R., 2013. Size-dependent piezoelectricity. Int. J. Solids Struct. 50(18), 2781-2791.

Hadjesfandiari A. R. 2014. Size-dependent thermoelasticity. Lat. Am. J. Solids Stru. 2014, 11(9), 1679-1708.

Hadjesfandiari, A. R., Dargush, G. F., 2011. Couple stress theory for solids. Int. J. Solids Struct. 48 (18), 2496-2510.

Hadjesfandiari, A. R., Dargush, G. F., 2015a. Evolution of generalized couple-stress continuum theories: a critical analysis. Preprint arXiv: 1501.03112.

Hadjesfandiari, A. R., Dargush, G. F., 2015b. Foundations of consistent couple stress theory. Preprint arXiv: 1509.06299.

Hadjesfandiari, A. R., Dargush, G. F., 2016. Couple stress theories: Theoretical underpinnings and practical aspects from a new energy perspective. Preprint arXiv: 1611.10249.

Hadjesfandiari, A. R., Hajesfandiari, A., Dargush, G. F., 2015. Skew-symmetric couple-stress fluid mechanics. Acta Mech. 226 (3), 871–895.

Kleman, M., Friedel, J., 2008. Disclinations, dislocations, and continuous defects: a reappraisal. Rev. Mod. Phys. 80, 61-115.

Kleman, M., Lavrentovich, O. D., 2003, Soft Matter Physics, an Introduction. Springer, New York.

Koiter, W. T., 1964. Couple stresses in the theory of elasticity, I and II. Proc. Ned. Akad. Wet. (B) 67, 17-44.

Kondo, K., 1968. On the two main currents of the geometrical theory of imperfect continua. In Mechanics of Generalized Continua, 200-213. Springer, Berlin, Heidelberg.

Kröner, E., 2001. Benefits and shortcomings of the continuous theory of dislocations. Int. J. Solids Struct. 38 (6-7), 1115-1134.

Kröner, E., 1968. Interrelations between various branches of continuum mechanics. In Mechanics of Generalized Continua, 330-340. Springer, Berlin, Heidelberg.

Kröner, E., 1958. Kontinuumstheorie der Versetzungen und Eigenspannungen. volume 5 of Ergebnisse der Angewandten Mathematik. Springer, Berlin, 7, 7-5.

Kröner, E. and Anthony, K.H., 1975. Dislocations and disclinations in material structures: The basic topological concepts. Annu. Rev. Mater. Sci., 5(1), 43-72.

Love, A.E.H., 1920. A treatise on the mathematical theory of elasticity. Cambridge University Press. 3$^{rd}$ Ed.

Lubliner, J., 2008. Plasticity theory. Courier Corporation.

Mindlin, R. D., Tiersten, H. F., 1962. Effects of couple-stresses in linear elasticity. Arch. Rational Mech. Anal. 11, 415–488.

Moore, M.R. and Hills, D.A., 2018. Solution of half-plane contact problems by distributing climb dislocations. Int. J. Solids Struct. 147, 61-66.

Nabarro, F.R.N., 1967. Dislocations in Solids. Oxford University Press, Oxford.

Schaefer, H., 1968. The basic affine connection in a Cosserat continuum. In Mechanics of Generalized Continua, 57-62. Springer, Berlin, Heidelberg.

Schaefer, H., 1967a. Die Spannungsfunktionen eines Kontinuums mit Momentspannungen, Bull. de l'Acad. Pol. Sci., Sér. Sci. Techn., I, 15, 1, 63; II, 15, 1.

Schaefer, H., 1967b. Das cosserat kontinuum. ZAMM-Journal of Applied Mathematics and Mechanics/Zeitschrift für Angewandte Mathematik und Mechanik, 47(8), 485-498.

Volterra, V., 1907. Sur l'équilibre des corps élastiques multiplement connexes. Ann. Sci. Ecol. Norm. Sup. III 24, 401–517.

Weingarten, J.,1901. Sulle superfici di discontinuita nella teoria dell'elasticita dei corpi solidi. Atti. Accad. Naz. Lincei, Cl. Sci. Fis. Mat. Natur. Rend.10(1), 57–60.